\documentclass[aps,pra,onecolumn,tightenlines,floats,superscriptaddress,11pt]{revtex4}

\usepackage{graphicx,braket,dsfont}
\usepackage{dcolumn,tikz}
\usepackage{bm}
\usepackage{color}
\usepackage{amssymb}
\usepackage{ams math}

\usepackage{hhline}
\usepackage{multirow}

\def\beq{\begin{equation}}
\def\eeq{\end{equation}}
\def\beqa{\begin{eqnarray}}
\def\eeqa{\end{eqnarray}}

\newcommand{\comm}[2]{\left[#1,#2\right]}

\newcommand*{\CparA}{\raisebox{-.2\height}{ \includegraphics[scale=0.2]{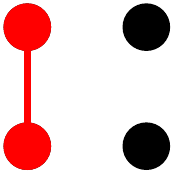}} }
\newcommand*{\CparB}{\raisebox{-.2\height}{ \includegraphics[scale=0.2]{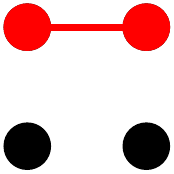}} }
\newcommand*{\CparC}{\raisebox{-.2\height}{ \includegraphics[scale=0.2]{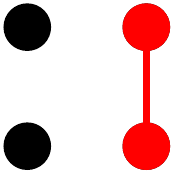}} }
\newcommand*{\CparD}{\raisebox{-.2\height}{ \includegraphics[scale=0.2]{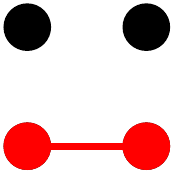}} }

\newcommand*{\CdisA}{\raisebox{-.2\height}{ \includegraphics[scale=0.2]{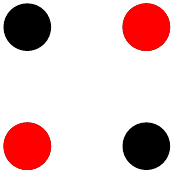}} }
\newcommand*{\CdisB}{\raisebox{-.2\height}{ \includegraphics[scale=0.2]{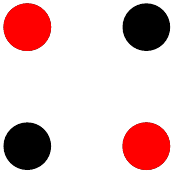}} }

\begin{document}
%%%%%%%%%%%%%%%%%%%%%%%%%%%%%%%%%%%%%%
%%%%%%%%%%%%%%%%%%%%%%%%%

%%%%%%%%%%%%%%%%%%%%%%%%%
% ARTICLE style
%\title{Effective spin physics in two-dimensional cavity QED arrays}
%\author{Ji\v{r}\'{i} Min\'{a}\v{r}, \c{S}ebnem G\"{u}ne\c{s} S\"{o}yler, Pietro Rotondo, Igor Lesanovsky}
%\affil[1]{School of Physics and Astronomy, The University of Nottingham, Nottingham, NG7 2RD, United Kingdom}
%\affil[2]{Centre for the Theoretical Physics and Mathematics of Quantum Non-equilibrium Systems, The University of Nottingham, Nottingham, NG7 2RD, United Kingdom}

% PRA style
\title{Effective spin physics in two-dimensional cavity QED arrays}
\author{Ji\v{r}\'{i} Min\'{a}\v{r}}
\affiliation{School of Physics and Astronomy, The University of Nottingham, Nottingham, NG7 2RD, United Kingdom}
\affiliation{Centre for the Theoretical Physics and Mathematics of Quantum Non-equilibrium Systems, The University of Nottingham, Nottingham, NG7 2RD, United Kingdom}
\author{\c{S}ebnem G\"{u}ne\c{s} S\"{o}yler}
\affiliation{School of Physics and Astronomy, The University of Nottingham, Nottingham, NG7 2RD, United Kingdom}
\affiliation{Centre for the Theoretical Physics and Mathematics of Quantum Non-equilibrium Systems, The University of Nottingham, Nottingham, NG7 2RD, United Kingdom}
\author{Pietro Rotondo}
\affiliation{School of Physics and Astronomy, The University of Nottingham, Nottingham, NG7 2RD, United Kingdom}
\affiliation{Centre for the Theoretical Physics and Mathematics of Quantum Non-equilibrium Systems, The University of Nottingham, Nottingham, NG7 2RD, United Kingdom}
\author{Igor Lesanovsky}
\affiliation{School of Physics and Astronomy, The University of Nottingham, Nottingham, NG7 2RD, United Kingdom}
\affiliation{Centre for the Theoretical Physics and Mathematics of Quantum Non-equilibrium Systems, The University of Nottingham, Nottingham, NG7 2RD, United Kingdom}

%%%%%%%%%%%%%%%%%%%%%%%%%%%%%%%%%%%%%%

\begin{abstract}
We investigate a strongly correlated system of light and matter in two-dimensional cavity arrays. We formulate a Jaynes-Cummings Hamiltonian for two-level atoms coupled to cavity modes and driven by an external laser field which reduces to an effective spin Hamiltonian in the dispersive regime. In one dimension we provide exact analytical solution. In two dimensions, we perform mean-field study and large scale quantum Monte Carlo simulations of both the Jaynes-Cummings and the effective spin models. We discuss the phase diagram and the parameter regime which gives rise to frustrated interactions between the spins. We provide quantitative description of the phase transitions and correlation properties featured by the system and we discuss graph-theoretical properties of the ground states in terms of graph colorings using P\'{o}lya's enumeration theorem.
\end{abstract}

\maketitle

%*******************************************************************************************************************************************************************************************************************
%*******************************************************************************************************************************************************************************************************************
%*******************************************************************************************************************************************************************************************************************

\section{Introduction}

Strongly coupled light-matter systems are at the heart of much of the effort in modern atomic and optical physics with applications ranging from quantum information processing to quantum simulations.

In this context, the use of cavities plays a prominent role as the strong confinement of the electromagnetic field implies strong interaction with matter coupled to the cavity modes. In particular, it offers possibilities to realize and study a plethora of quantum light-matter many-body Hamiltonians such as the so-called Jaynes-Cummings-Hubbard or Rabi-Hubbard models \cite{Koch_2009,Schmidt_2009,Bhaseen_2009,Mering_2009,Silver_2010,Nunnenkamp_2011,Ciccarello_2011,Tan_2011,Mascarenhas_2012,Quach_2013,Zhu_2013}, or quantum fluids of light, where the effective interaction between light fields is mediated by a non-linear medium \cite{Carusotto_2013,Bliokh_2015b,Bliokh_2015}. This offers ways to study various physical phenomena such as excitation propagation in chiral networks \cite{Petersen_2014,Pichler_2015, Ramos_2016}, the physics of spin glasses \cite{Strack_2011, Buchhold_2013, Rotondo_2015b}, self-organization of the atomic motion in optical cavities \cite{Domokos_2002,Zippilli_2004,Asboth_2005,Black_2003,Baumann_2010} or quantum phase transitions in arrays of nanocavity quantum dots \cite{Grochol_2009} and in Coulomb crystals \cite{Bermudez_2012}. Furthermore, modern implementations of optical and microwave cavities allow for the creation of lattices with tunable geometries and dimensionality \cite{Hartmann_2008a,Tomadin_2010,Schmidt_2013}.

The paradigmatic description of cavity and circuit QED systems is typically in terms of the famous Jaynes-Cummings (JC) \cite{Jaynes_1963} or Dicke \cite{Dicke_1954} models, which describe the interaction between the modes of the light field and the matter constituents, typically spin or phononic degrees of freedom of atoms or ions. Importantly, effective spin models emerge in the dispersive limit of the JC or Dicke models \cite{Porras_2004,Deng_2005}. Under some circumstances this leads to spin Hamiltonians with frustrated or long-range interactions \cite{Kim_2009, Lin_2011}, thus offering ways to study rich physics of quantum magnetism. This is a particularly interesting direction allowing e.g. for studies of spin liquids \cite{Balents_2010, Savary_2016, Zhou_2016} with optical quantum simulators.

While advanced numerical techniques, such as tensor networks, have been developed for spin Hamiltonians \cite{Schollwock_2005,Verstraete_2008,Schollwock_2011,Orus_2014}, the use of similarly efficient techniques for quantum optical systems, where a system of spins is coupled to the bosonic modes of an electromagnetic field remains limited. In this work we use mean-field (MF) description, exact diagonalization and large-scale quantum Monte Carlo (QMC) algorithm to study arrays of waveguide cavities (we note that in the context of cavity QED, QMC was implemented to study both the Jaynes-Cummings-Hubbard \cite{Hohenadler_2011} and the Rabi-Hubbard \cite{Flottat_2016} models). This gives us rigorous tools to investigate the emergence of the effective spin physics as a limiting case of the parent JC Hamiltonian for arbitrary lattice geometries and dimensions. Specifically, in this work we focus on square lattice geometry and we study the ground state properties of the JC and the effective spin models for various parameter regimes. We then show, that depending on the parameter regime and the array geometry, spin models with both non-frustrated and frustrated interactions can be engineered.

The paper is organized as follows. In Sec. \ref{sec:The model} we introduce the system and derive the effective spin model from the parent JC Hamiltonian. In Sec. \ref{sec:1D} we present exact analytical solution of the spin model in one dimension. In Sec. \ref{sec:Simulations} we present the results of the QMC simulation and MF analysis and discuss different regimes provided by the investigated model. We explain how the present work opens possibilities for simulating frustrated systems in Sec. \ref{sec:Frustration} and conclude in Sec. \ref{sec:Conclusions}.

%****************************************************************************************************************************************************************************************************************************
%****************************************************************************************************************************************************************************************************************************
\begin{center}
\begin{figure}[h!]
%\vspace{-2.0cm}
\includegraphics[width=0.9\columnwidth]{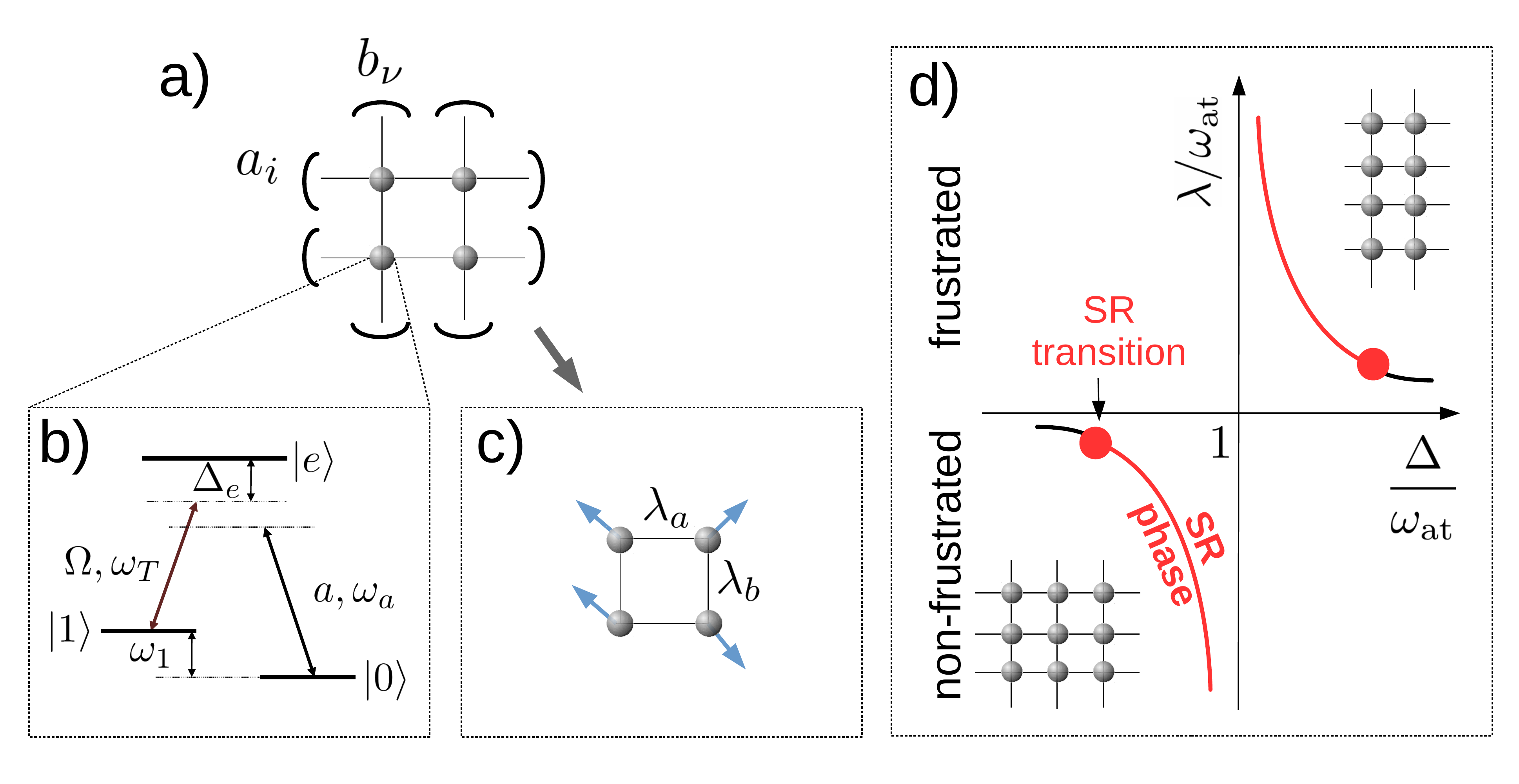}
%\vspace{-1.cm}
\caption{Sketch of the considered coupled light-matter system. (a) Horizontal (vertical) cavity modes $a$ ($b$) couple to three-level systems located at the nodes of the array and represented by the grey spheres. (b) The atomic level scheme. The $\ket{1}-\ket{e}$ transition is driven by a classical field with Rabi frequency $\Omega$ and the excited state $\ket{e}$ is adiabatically eliminated, see text for details. (c) Schematic of the emergent spin system after the elimination of the cavity modes. (d) Graphical representation of the parameter regime of the emergent spin Hamiltonian (\ref{eq:H spin main}). The sign of the effective spin-spin interaction $\lambda$ determines the nature of the spin-spin interaction: non-frustrated if $\lambda<0$, frustrated if $\lambda>0$. As $|\lambda|$ is increased, a transition to a superradiant (SR) phase occurs, corresponding to a non-zero spin excitations of the system. While arbitrary rectangular arrays can be considered in the non-frustrated regime, only elongated geometries give rise to a non-trivial spin physics in the frustrated regime, cf. Sec. \ref{sec:Frustration} for details.} 
%\vspace{-2.cm}
\label{fig:scheme}
\end{figure}
\end{center}

%****************************************************************************************************************************************************************************************************************************
\section{Effective Hamiltonians}
\label{sec:The model}

\subsection{Jaynes-Cummings Hamiltonian}
\label{sec:JC Ham}

Recent advances in integrated optical circuits, where in principle arbitrary waveguide geometries can be created with high precision by laser engraving in the silica substrate \cite{Marshall_2009} and an active experimental effort to combine the waveguides with atomic microtraps on a single device \cite{Derntl_2014, Potts_2016} motivate us to investigate a system of three-level atoms embeded in waveguide cavities. 

We consider a square cavity array, where we denote by $a_i$ ($b_\nu$) the modes in the $i$-th row ($\nu$-th column) of the array, as shown in Fig. \ref{fig:scheme}a. We use the latin (greek) indices to denote the rows (columns) throughout the article. All sites of the array are occupied by identical three-level systems in a $\Lambda$ configuration, where $\ket{0},\ket{1}$ denote the ground states and $\ket{e}$ the excited state, see Fig. \ref{fig:scheme}b. The cavity modes are coupled with strength $g_0$ to the $\ket{0}-\ket{e}$ transition, while the $\ket{1}-\ket{e}$ transition couples to a classical field $\Omega$ with frequency $\omega_T$ which propagates perpendicularly to the plane of the array and which is detuned by $\Delta_e = \omega_{1} - \omega_T$ with respect to the $\ket{1}-\ket{e}$ transition. In the limit $\Delta_e \gg g_0,\Omega$ one can eliminate the excited state, which we described in detail in our previous publication \cite{Minar_2016_pub}. Furthermore, under the condition of strong driving
\beq
  \Omega \gg g_0
  \label{eq:cond Omega g}
\eeq
the resulting Hamiltonian is of the JC type and reads (see Appendix \ref{app:Adiabatic Elimination} for the details of the derivation and of the full Hamiltonian)
\begin{subequations}
\begin{align}
H_{\rm JC}=&\sum_i\Delta_ia^{\dag}_ia_i+\sum_{\nu}\Delta_{\nu}b^{\dag}_{\nu}b_{\nu}+\frac{\omega_{\rm at}}{2}\sum_{i,\nu}\sigma^z_{i\nu}+ \nonumber \\
      &g \sum_{i,\nu} \left( \sigma^+_{i\nu}a_{i}+a^{\dag}_{i}\sigma^-_{i\nu} \right)+g \sum_{i,\nu} \left( \sigma^+_{i\nu}b_{\nu}+b^{\dag}_{\nu}\sigma^-_{i\nu} \right).      
\end{align}
\label{eq:H JC}
\end{subequations}
with the effective atomic transition frequency and the effective coupling strength
\begin{subequations}
\begin{align}
     \omega_{\rm at} &= - \frac{\Omega^2}{\Delta_e} \label{eq:pars H JC omega} \\ 
     g &= - \frac{g_0 \Omega}{\Delta_e}.     \label{eq:pars H JC g}
\end{align}
\end{subequations}
Here, $\sigma^{z,\pm}_{i\nu}$ are the usual Pauli matrices written in the $\{\ket{1},\ket{0}\}$ basis, $\Delta_{i(\nu)}$ the effective cavity frequencies, $\omega_{\rm at}$ the effective frequency of the $\ket{0}-\ket{1}$ transition and $i=1..L_y, \nu=1..L_x$, where $L_y (L_x)$ is the number of rows (columns) of the array.

%****************************************************************************************************************************************************************************************************************************
\subsection{Spin Hamiltonian}
\label{sec:spin Ham}

In the large detuning limit 
\beq
    g \ll |\omega_{\rm at} - \Delta_{i(\nu)}|,
    \label{eq:cond elimination}
\eeq
one can further eliminate the cavity fields to obtain an effective spin Hamiltonian, where a given spin is coupled to all other spins belonging to the same cavity mode (see Fig. \ref{fig:scheme}c and Appendix \ref{app:Adiabatic Elimination}),
\begin{equation}
 H_{\rm spin}= H_{\rm spin,0} + \delta H + H_{\rm spin,int}
\label{eq:H spin main}
\end{equation}
where
\begin{subequations}
 \begin{align}
 H_{\rm spin,0} &= \sum_{i,\nu} \left( \frac{\omega_{\rm at}}{2} + \lambda_i + \lambda_\nu \right) \sigma^z_{i\nu} \label{eq:H spin 0}\\
 \delta H &= \sum_{i,\nu} \delta \omega_{{\rm at},i\nu}\sigma^z_{i\nu} \label{eq:delta H}\\
 H_{\rm spin,int} &= \sum_{i,\nu \neq \mu}\lambda_{i}(\sigma^-_{i\nu}\sigma^+_{i\mu}+\sigma^+_{i\nu}\sigma^-_{i\mu})+\sum_{i \neq j,\nu}\lambda_{\nu}(\sigma^-_{i\nu}\sigma^+_{j\nu}+\sigma^+_{i\nu}\sigma^-_{j\nu}), \label{eq:H spin int}
\end{align}
\end{subequations}
where
\begin{subequations}
 \begin{align}
  \lambda_i &= -\frac{g^2}{2 (\Delta_i - \omega_{\rm at})} \label{eq:lambda i} \\
  \lambda_\nu &= -\frac{g^2}{2 (\Delta_\nu - \omega_{\rm at})} \label{eq:lambda nu}
 \end{align}
 \label{eq:pars spin main}
\end{subequations}
are the effective spin-spin couplings along the rows (columns) and 
\beq
  \delta \omega_{{\rm at},i\nu} = \lambda_i \left( 2 a^\dag_i a_i + a^\dag_i b_\nu + b^\dag_\nu a_i \right) + \lambda_\nu \left( 2 b^\dag_\nu b_\nu + a^\dag_i b_\nu + b^\dag_\nu a_i \right).
  \label{eq:delta omega spin}
\eeq

%****************************************************************************************************************************************************************************************************************************
\subsection{Validity of the approximations and frustrated vs. non-frustrated regime}
\label{sec:regimes}

The elimination of photonic or phononic fields giving rise to an effective spin physics is a known technique often used in the design of various quantum optical simulators. It can lead to interesting frustrated spin Hamiltonians, e.g. in the context of trapped ions \cite{Kim_2009, Lin_2011, Grass_2016}.

In order to simplify the parameter space, in what follows we choose all the couplings to be the same along rows (columns): $\lambda_a \equiv \lambda_i,\; \forall i$ ($\lambda_b \equiv \lambda_\nu,\; \forall \nu$). Schematically, the parameter regime of the spin Hamiltonian (\ref{eq:H spin main}) is summarized in Fig. \ref{fig:scheme}d.

First we note, that the parameters $\omega_{\rm at}/2$, $\lambda$ and $\delta \omega_{\rm at}$ of the Hamiltonian (\ref{eq:H spin main}) given by (\ref{eq:pars H JC omega}),(\ref{eq:pars spin main}) and (\ref{eq:delta omega spin}) can take both positive or negative sign. In particular the sign of $\lambda$ determines the kind of physical situation provided by the interaction Hamiltonian (\ref{eq:H spin int}): \emph{non-frustrated} if both couplings are negative, $\lambda_{a(b)} <0$ and \emph{frustrated} otherwise. This is apparent from the form of the interaction which tends to align each pair of spins antiparallel whenever the corresponding coupling is positive. This then leads to frustration as the antiparallel alignment cannot be satisfied simultaneously for all the spins. Note that while we consider square lattice for concreteness, the presence of frustration in cavity arrays stems from all-to-all interaction between spins belonging to the same cavity mode and, hence, is independent of the lattice geometry.

Next, we discuss the parameter regimes of the Hamiltonian (\ref{eq:H spin main}). We recall that the only requirement used in the derivation of (\ref{eq:H spin main}) from the parent JC Hamiltonian (\ref{eq:H JC}) is the condition (\ref{eq:cond elimination}), $g \ll |\omega_{\rm at} - \Delta_{a(b)}|$.

\emph{(i) Weakly interacting regime.}
We refer to the weakly interacting regime as the regime where (we drop the $a,b$ indices for simplicity)
\beq
	|\lambda| \ll |\omega_{\rm at}|.
	\label{eq:cond 3 main}
\eeq
Here, we have neglected the $\delta \omega_{\rm at}$ term contributing to the atomic transition frequency since $\delta \omega_{\rm at} \propto \lambda$ \footnote{One should verify the self-consistency of the condition (\ref{eq:cond 3 main}) when performing the simulation of the parent JC model, i.e. to check, whether the resulting cavity occupation is such that $|\braket{\delta \omega_{\rm at}}| \ll |\omega_{\rm at}|$.}. One should verify that reaching the weakly interacting regime is compatible with the conditions (\ref{eq:cond Omega g}), (\ref{eq:cond elimination}) used in the derivation of the Hamiltonians (\ref{eq:H JC}), (\ref{eq:H spin main}). It is easy to show that it is indeed the case: substituting (\ref{eq:pars spin main}) for $\lambda$ in (\ref{eq:cond 3 main}), we get $|g^2/\omega_{\rm at}| \ll |\Delta - \omega_{\rm at}|$. This implies, together with (\ref{eq:cond elimination}), that $|g| \lesssim |\omega_{\rm at}|$. Finally, substituting for $g$ from (\ref{eq:pars H JC g}), we get $|g_0| \lesssim |\Omega|$. This is enforced by the stronger condition (\ref{eq:cond Omega g}), which completes our consistency check.

\emph{(ii) Strongly interacting regime.} Here we refer to the strongly interacting regime as the regime where $|\lambda| \gtrsim |\omega_{\rm at}/2 + \braket{\delta \omega_{\rm at}}|$. Here, the cavity fields dependence of the $\delta \omega_{\rm at}$ term plays an essential role. We leave this interesting possibility for Sec. \ref{sec:Frustration} and focus first on the scenario where the dynamics of the cavity fields decouples from the spins leading to a pure spin Hamiltonian.

%****************************************************************************************************************************************************************************************************************************
%****************************************************************************************************************************************************************************************************************************
\section{1D: Exact solution of the spin model}
\label{sec:1D}

It is illustrative to clarify on a simple example some of the basic properties of the Hamiltonian (\ref{eq:H spin main}). Specifically, we are interested in the nature of phase transitions featured by (\ref{eq:H spin main}) and the scalings of critical couplings. To this end we consider a one dimensional limit of (\ref{eq:H spin main}) by taking a single cavity mode $a$. The Hamiltonian simplifies to
\beq
	H_{\rm spin}^{\rm 1D} = \Delta a^\dag a + \left[ \omega_{\rm at} + 4 \lambda a^\dag a \right] J^z + 2 \lambda J^+ J^-,
	\label{eq:H spin 1D}
\eeq
Here, $J$ are the total angular momentum operators
\begin{subequations}
\begin{align}
	J^l &= \frac{1}{2} \sum_{i=1}^N \sigma^l_i, \;  \; l=x,y,z\\
	J^\pm &= \sum_{i=1}^N \sigma^\pm_i.
\end{align}
\end{subequations}
We note that in the absence of the cavity fields, (\ref{eq:H spin 1D}) is the well-known Lipkin-Meshkov-Glick model \cite{Lipkin_1965}, which has been recently studied also in the context of cavity QED \cite{Morrison_2008}. The advantage of the model (\ref{eq:H spin 1D}) is that it is exactly solvable providing us with useful analytical insights. Using the usual angular momentum algebra
\begin{subequations}
\begin{align}	
	%\comm{J^z}{J^\pm} &=& \pm J^\pm \\
	%\comm{J^+}{J^-} &=& 2 J^z \\
	J^z \ket{J,m} &= m\ket{J,m} \\
	J^\pm \ket{J,m} &= \sqrt{J(J+1)-m(m\pm 1)} \ket{J,m \pm 1}, \label{eq:angular momentum stop}
\end{align}
\end{subequations}
where $J$ is the half-integer total angular momentum ($J=N/2$ for $N$ spins) and $m = -J,-J+1,...,J$, it follows that $\ket{J,m,n}$, where $n$ is the photon number, are the eigenstates of the Hamiltonian (\ref{eq:H spin 1D}). The eigenenergies are
\beqa
	E_{J,m,n} &=& \Delta n + \left[ \omega_{\rm at} + 4 \lambda n \right] m + 2 \lambda \left[ J(J+1) - m(m-1) \right] \nonumber \\
	&=& \left[ \Delta + 4 \lambda m \right] n + 2 \lambda \left[ J(J+1) - m(m-1) \right] + \omega_{\rm at} m,
	\label{eq:E_Jmn}
\eeqa
where in the second line we have regrouped the terms in order to emphasize the dependence on the photon number $n$. 

The implications of the first bracket in the second line of (\ref{eq:E_Jmn}) are the following. For
\beq
	E_n \equiv \Delta + 4 \lambda m > 0
	\label{eq:E_n}
\eeq
the ground state photon number is 0. On the other hand, for $E_n < 0$ the ground state photon number is $n=\infty$, which invalidates the approximate description in terms of the effective Hamiltonian (\ref{eq:H spin 1D}). At this point it is important to note that since both $\Delta-\omega_{\rm at}$ in the denominator of $\lambda$ and $m$ can take positive or negative values, there is always a combination of $m$ and $\Delta-\omega_{\rm at}$ where the transition $n=0 \leftrightarrow n=\infty$ occurs as $\lambda$ is varied. The situation is summarized in Table \ref{tab:transitions}. The main message contained in the Table \ref{tab:transitions} is that it is \emph{not} possible to simulate the frustrated spin system using (\ref{eq:H JC}) in one dimension (see also \cite{Rotondo_2015}). In Sec. \ref{sec:Frustration} we will show, how this limitation can be circumvented in two dimensions by exploiting the properties of the $\delta \omega_{\rm at}$ term (\ref{eq:delta omega spin}).

In what follows we shall investigate this transition and its relation to the parent JC Hamiltonian (\ref{eq:H JC}) further. The $n=0 \leftrightarrow n=\infty$ transition occurs when $E_n$ changes sign. From (\ref{eq:pars spin main}) and (\ref{eq:E_n}) we get the expression for a critical coupling $g_c$
\beq
  g_c = \sqrt{\frac{\Delta(\Delta - \omega_{\rm at})}{2 m}}.
\eeq
Lets first consider $\omega_{\rm at} > 0$. In the non-frustrated case ($\lambda<0$, $\Delta > \omega_{\rm at}$), the minimal possible value of $g_c$ corresponds to $m=N/2$ (i.e. all spins excited). On the other hand, for $\lambda>0$ and positive $\omega_{\rm at}$ assumed here, we can have either $\Delta>0$ or $\Delta<0$. For $\Delta<0$, we can see immediately from (\ref{eq:E_n}) that $E_n$ can be made \emph{always} negative by a suitable choice of $m$. Specifically, considering the spin ground state $m=-N/2$, the global ground state would correspond to $n=\infty$ invalidating the description in terms of (\ref{eq:H spin 1D}). On the other hand, for $\Delta>0$ the system undergoes the $n=0 \leftrightarrow n=\infty$ transition as $\lambda$ is increased. However, it occurs for $m=-N/2$, i.e. before any spin transition could possibly take place. One could now proceed analogously for $\omega_{\rm at}<0$ \footnote{We note that (\ref{eq:H spin 1D}) does not inherit the simple $\mathbb{Z}_2$ symmetry of the parent JC model, namely the symmetry under simultaneous exchange $\sigma^z \rightarrow -\sigma^z, \sigma^{+}a \rightarrow a^\dag \sigma^-$ and $\omega_{\rm at} \rightarrow -\omega_{\rm at}$, due to the non-linear nature of the transformation yielding the effective spin model, cf. Appendix \ref{app:Adiabatic Elimination}.}. 

In summary, the critical coupling at which the $n=0 \leftrightarrow n=\infty$ transition occurs is given by
\beq
	g^{\rm ph}_c = \frac{1}{\sqrt{N}} \sqrt{\Delta(\Delta - \omega_{\rm at})},
	\label{eq:g crit ph}
\eeq
where we have emphasized by the label "ph", that the transition is in the photon number. 

In one dimension, the only non-trivial situation is thus the non-frustrated case, $\lambda<0$, where $n=0$. Here, a series of transitions between phases with $N_{\rm exc}(m)$ and $N_{\rm exc}(m+1) = N_{\rm exc}(m)+1$ excited spins takes place as $|\lambda|$ is increased (here $N_{\rm exc} = (2m+N)/2$). The corresponding coupling strengths at which these transitions occur are obtained from (\ref{eq:E_Jmn}) by solving for $E_{J,m,0} = E_{J,m+1,0}$. 

For instance, considering $\omega_{\rm at}>0$ and $\Delta > \omega_{\rm at}$, third line in Table \ref{tab:transitions}, the first spin transition from $m=-N/2$ to $m=-N/2+1$ occurs at
\beq
	g_c = \frac{1}{\sqrt{N}} \sqrt{\omega_{\rm at} (\Delta - \omega_{\rm at})}.
	\label{eq:g crit spin 1D}
\eeq
One can also read off from the expression (\ref{eq:g crit spin 1D}) the scaling properties of the critical point of the spin transition with the system size, $g_c \propto 1/\sqrt{N}$ and correspondingly for the critical coupling $\lambda$, $|\lambda_c| \propto 1/N$.

\begin{table}
\begin{tabular}{|c|c|c|c|c|c|}
  \hline			
   $\omega_{\rm at}$ & \parbox[c]{2cm}{ground state \\ configuration} & $\lambda$ & $\Delta - \omega_{\rm at}$ & $\Delta$ & spin transition \\
   \hhline{|=|=|=|=|=|=|}
  \multirow{3}{*}{+} & \multirow{3}{*}{$\ket{g..g}$} & \multirow{2}{*}{+} & \multirow{2}{*}{--} & + & no \\ \cline{5-6}
  &  &  &  & -- & no ($n = \infty$) \\  \cline{3-6}
  & & -- & + & + & yes \\  
  \hhline{|=|=|=|=|=|=|}
  \multirow{3}{*}{--} & \multirow{3}{*}{$\ket{s..s}$} & \multirow{2}{*}{--} & \multirow{2}{*}{+} & -- & yes \\ \cline{5-6}
  &  &  &  & + & no ($n = \infty$) \\  \cline{3-6}
  & & + & -- & -- & no \\
  \hline  
\end{tabular}
\caption{Summary of phase transitions in the 1D effective spin model (\ref{eq:H spin 1D}). $+$ ($-$) stand for positive (negative) values respectively. The ``$(n=\infty)$'' description in the second and fifth line indicates that in these cases, the ground state corresponds to the infinite photon number independently of the value of $\lambda$, see (\ref{eq:E_n}) and text for details.}
\label{tab:transitions}
\end{table}

%****************************************************************************************************************************************************************************************************************************
%****************************************************************************************************************************************************************************************************************************
\section{2D: Analytical study and QMC simulations}
\label{sec:Simulations}

%****************************************************************************************************************************************************************************************************************************
%****************************************************************************************************************************************************************************************************************************
After having analyzed the situation in 1D, we now turn our attention to 2D. It is well known that the JC model features second order superradiant phase transition as the coupling strength is varied \cite{Dicke_1954, Hepp_1973, Wang_1973}. We will analyse the scaling properties at this superradiant phase transition and evaluate the two-point spin correlation functions of the cavity array. In order to do so, we employ large scale QMC simulations using the worm algorithm \cite{Prokofjev_1998a,Prokofjev_1998b}. In the following we compare the QMC results with the MF solutions. We emphasize that in the considered square lattice geometry the spins are \emph{not} all-to-all connected (they are connected only if they belong to the same row/column), i.e. it is not apriori obvious whether the MF solutions provide an accurate quantitative description.

In order to simplify the discussion, in this section we consider equal couplings along all rows and columns, $\lambda \equiv \lambda_a = \lambda_b$ (i.e. $\Delta \equiv \Delta_i = \Delta_\nu,\; \forall i,\nu$). Motivated by the findings in the one-dimensional case, we focus only on the non-frustrated case with $\omega_{\rm at}>0$ and $\Delta>\omega_{\rm at}$. We will address the frustrated case in Sec. \ref{sec:Frustration}.

\emph{Mean-field solutions. } In the thermodynamic limit, one can find MF solutions of the JC model (\ref{eq:H JC}) which we describe in detail in Appendix \ref{app:MF} and which we use for the sake of comparison with the QMC data. In particular, for an array of size $N=L_x \times L_y$ one can find expressions for the critical strength $g_c^{\rm MF}$ of the coupling at which the superradiant transition occurs and the number of spin excitations $N_{\rm exc}$ in the superradiant phase, which read
\beq
  g_c^{\rm MF} = \sqrt{\Delta \omega_{\rm at} \left( \frac{L_x+L_y}{4 L_x L_y} \right)}
  \label{eq:gc MF}
\eeq
and
\beq
  N_{\rm exc}^{\rm MF} = \frac{N}{2} \left(1- \left( \frac{g_c^{\rm MF}}{g} \right)^2 \right)
  \label{eq:Nexc MF}
\eeq
respectively. In the specific case of a square array $L \equiv L_x = L_y$ and in the limit $\Delta \rightarrow \infty$, where the descriptions in terms of (\ref{eq:H JC}) and (\ref{eq:H spin main}) should coincide, we get with the help of (\ref{eq:pars spin main})
\beq
  \lambda_c^{\rm MF,\infty} \equiv -\frac{\omega_{\rm at}}{4 L}
  \label{eq:lambda MF inf}
\eeq
and
\beq
  N_{\rm exc}^{{\rm MF},\infty} \equiv \frac{N}{2} \left(1 + \frac{1}{4 L}\frac{\omega_{\rm at}}{\lambda} \right).
  \label{eq:Nexc MF inf}
\eeq
~\\
~\\
\begin{center} 
\begin{figure}[h!]
%\vspace{-2.0cm}
\includegraphics[width=\columnwidth]{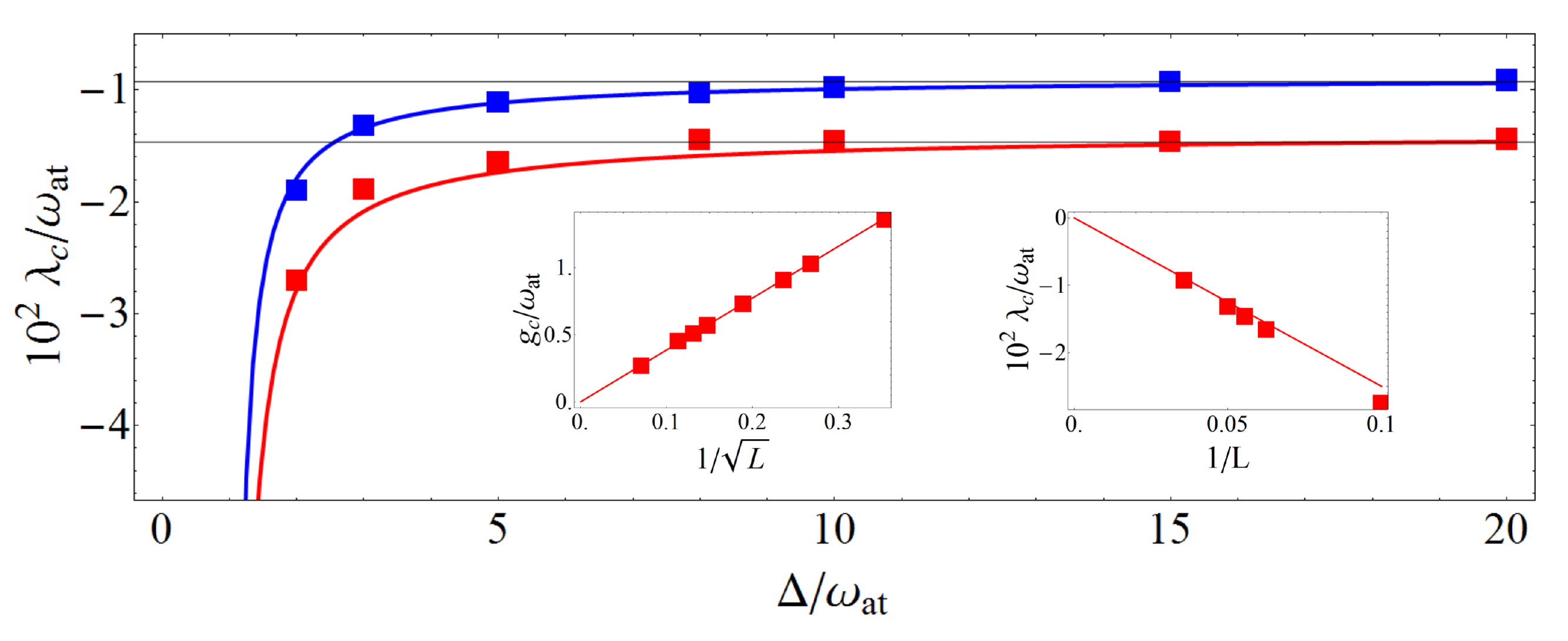}
%\vspace{-1.cm}
\caption{Reaching the effective spin model as the limiting case of the JC Hamiltonian. The main plot shows critical coupling $\lambda_c$ of the superradiant phase transition as a function of the photon detuning. The squares are the data obtained from the QMC simulation of the JC model (\ref{eq:H JC}). The solid lines are the MF predictions (\ref{eq:gc MF}). The red (blue) data correspond to system sizes $N=L \times L = 18 \times 18$ ($N=28 \times 28$) respectively. The solid black lines are the critical coupling values obtained from the QMC simulation of the spin model (\ref{eq:H spin main}) for a given system size. \emph{Left inset}: Finite size scaling of the critical coupling of the JC model. \emph{Right inset}: Finite size scaling of the critical coupling of the spin model. We note that the coupling goes to zero in the thermodynamics limit as expected, cf. Eq. (\ref{eq:gc MF}). We have used $\Delta/\omega_{\rm at}=30$ in the insets.}
%\vspace{-2.cm}
\label{fig:crossover}
\end{figure}
\end{center}
The spin model (\ref{eq:H spin main}) is an effective description of the parent JC model (\ref{eq:H JC}) in the limit of large detuning (\ref{eq:cond elimination}). Therefore, the excitations of the JC model in the superradiant phase result in spin excitations in the effective spin model. Here, QMC provides an efficient numerical tool to study this limit behaviour of the JC model and how well it is described by the effective spin model. The results of the simulations are presented in Fig. \ref{fig:crossover}. Here we show the critical couplings of the superradiant phase transition $g_c$ determined using QMC (using the total number of photonic and spin excitations as order parameter, square data points) and the MF prediction (\ref{eq:gc MF}), solid lines. The red (blue) data correspond to two different system sizes $N=18 \times 18$ ($N=28 \times 28$) respectively and the horizontal black lines are the values of the critical couplings $\lambda_c$ obtained from the QMC simulation of the effective spin model (\ref{eq:H spin main}). As expected, we find that the values of the critical couplings approach asymptotically in the limit $\Delta \gg \omega_{\rm at}$ where the two models (\ref{eq:H JC}) and (\ref{eq:H spin main}) coincide. In the insets we show the finite size scaling of the critical couplings for the JC (left inset) and effective spin (right inset) models. As in the main plot, the squares represent the QMC data and the solid lines are the MF predictions (\ref{eq:gc MF}) and (\ref{eq:lambda MF inf}). The slight departure of the scaling for the spin model for small system sizes is indeed a finite size effect on which we will comment momentarily. We also note, that the couplings for the present 2D model scale in the same way as the 1D predictions (\ref{eq:g crit spin 1D}), i.e. in the linear extent of the system, $g_c \propto 1/\sqrt{L}$. This is due to the fact that the scaling is determined by the number of \emph{cavity modes} to which the atoms couple rather than by the system size $N$ (see also Appendix \ref{app:MF}).

\begin{center} 
\begin{figure}[h!]
%\vspace{-2.0cm}
\includegraphics[width=0.8\columnwidth]{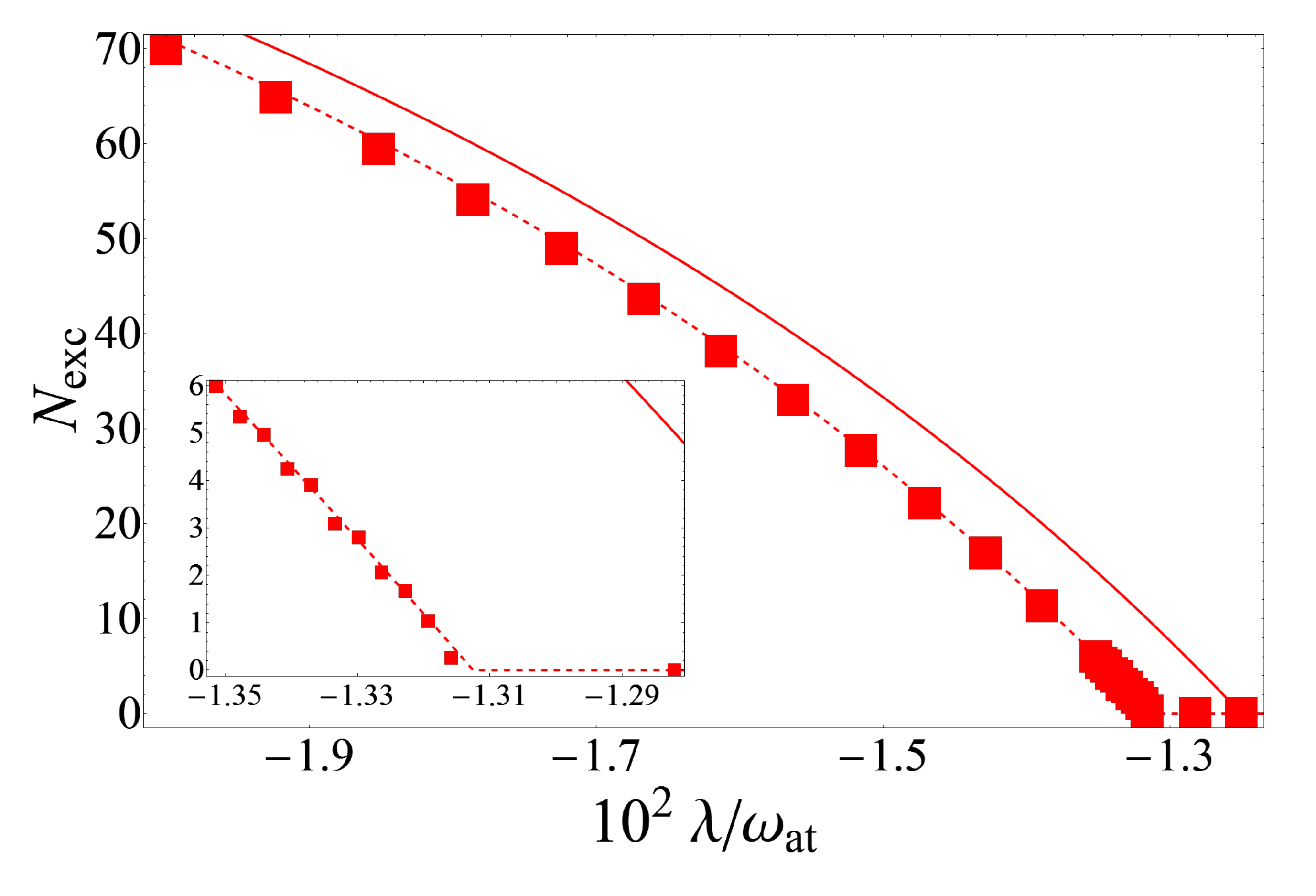}
%\vspace{-1.cm}
\caption{Number of spin excitations of the spin model (\ref{eq:H spin main}) on $N=20 \times 20$ array vs. the interaction strength $\lambda$. The squares represent the QMC data, the solid line is the MF prediction (\ref{eq:Nexc MF inf}). The dashed line is the MF prediction corrected for the finite size offset (see text for details). The inset shows the magnification of the region in the vicinity of the critical coupling.}
%\vspace{-2.cm}
\label{fig:exc}
\end{figure}
\end{center}

After we have verified, that the critical couplings of the JC model coincide with those of the spin model, we now study the number of excitations of the spin model as the coupling strength is varied. This is shown in Fig. \ref{fig:exc}. The solid line corresponds to the MF prediction (\ref{eq:Nexc MF inf}). The squares correspond to the QMC simulation of the spin model (\ref{eq:H spin main}) on a $N=20 \times 20$ array, where we have neglected $\lambda_a,\lambda_b$ in (\ref{eq:H spin 0}) as $|\lambda| \ll \omega_{\rm at}$ in the studied regime. The discrepancy between the MF prediction and the QMC simulation is precisely the consequence of neglecting $\lambda_a,\lambda_b$ in (\ref{eq:H spin 0}) and results in an offset $-1/(4 L^2)$ in the values of $\lambda$ - the dashed line corresponds to the MF solution corrected for this offset \footnote{In the spin model, the transition point is obtained from $E^{0\rm exc} = E^{1\rm exc}$, where $E^{0\rm exc} = -\frac{\omega_{\rm at}'}{2}N$ is the ground state energy with no spin excitation and $E^{1\rm exc} = \frac{\omega_{\rm at}'}{2}(-N+2)+4\lambda(L-1)$ the energy with a single excitation (here we have used (\ref{eq:Eigs M})). Solving for $\lambda$ we find $\lambda_c = -\frac{\omega_{\rm at}}{4 L}$ which coincides with the MF result (\ref{eq:lambda MF inf}). Neglecting the $\lambda$ terms in $\omega_{\rm at}'$ (\ref{eq:omega prime}) amounts to replacing $\omega_{\rm at}'$ by $\omega_{\rm at}$ in $E^{0\rm exc}=E^{1 \rm exc}$ above, yielding the solution $\lambda_c = -\frac{\omega_{\rm at}}{4(L-1)} \approx -\frac{\omega_{\rm at}}{4 L} - \frac{\omega_{\rm at}}{4 L^2}$, where the last term is the offset used in Fig. \ref{fig:exc}. }.

So far, we concentrated only on one-point observables in our QMC simulations and found a good agreement with the MF predictions. In order to go beyond the MF picture we next consider the correlation functions. Before presenting the results and in order to get a deeper insight in the structure of the spin Hamiltonians emergent in cavity arrays, in the following section we study the properties of the ground states from the group and graph theory perspective.

%****************************************************************************************************************************************************************************************************************************
%****************************************************************************************************************************************************************************************************************************
\subsection{Ground state structure}

\emph{Symmetry considerations.} We start our analysis in this section by noting that the total number of spin excitations, $\hat{N}_{\rm exc} = \sum_{i,\nu} \frac{1}{2} \left( \sigma^z_{i\nu} + 1 \right)$, is the constant of motion of the Hamiltonian (\ref{eq:H spin main}). This significantly simplifies the problem in that in order to find the ground states of (\ref{eq:H spin main}), one only needs to solve for the eigenstates of the interaction Hamiltonian (\ref{eq:H spin int})
\beq
	H_{\rm spin, int} \ket{\psi_{\rm GS}} = E_{\rm GS} \ket{\psi_{\rm GS}}
\eeq
in the given excitation number sector $N_{\rm exc}$.

The problem can be further simplified by exploiting the real space symmetries of the Hamiltonian (\ref{eq:H spin main}), similarly to the analysis carried out e.g. for the antiferromagnetic Heisenberg chain \cite{Ishino_1990}. Considering the most symmetric situation, i.e. a square array with equal couplings ($L_x = L_y$, $\lambda_a = \lambda_b$), the discrete symmetry group of the Hamiltonian (\ref{eq:H spin main}) is $G_{L_x \times L_x}=\{\mathfrak{R},\mathfrak{C}\} \cup D_4$, where $\mathfrak{R}, \mathfrak{C}$ and $D_4$ stand for permutation of rows, permutation of columns and the dihedral group of the square array (i.e. successive rotations $r_{\pi/2}, r_{\pi}, r_{3\pi/2}$ by $\pi/2$ and reflections about the horizontal ($h$), vertical ($v$) and the two diagonal($p,n$) axes of the array) respectively. In order to get use of the symmetries, one has to find the irreducible representations (irreps) of $G$. While a systematic approach exists for finding irreps of the full symmetric group $S_{N=L_x L_y}$ \cite{Hamermesh_1962}, the subduced representations of the subgroup $G \subset S_N$ are in general reducible \cite[ch.3]{Ma_2007}. Motivated by exact diagonalization results, instead of finding the irreps of $G$, we focus on the graph-theoretical properties of the ground states in what follows.

\begin{center}
	\begin{figure}[t!p]
	\hspace*{-0.5cm}
  	\includegraphics[width=13cm]{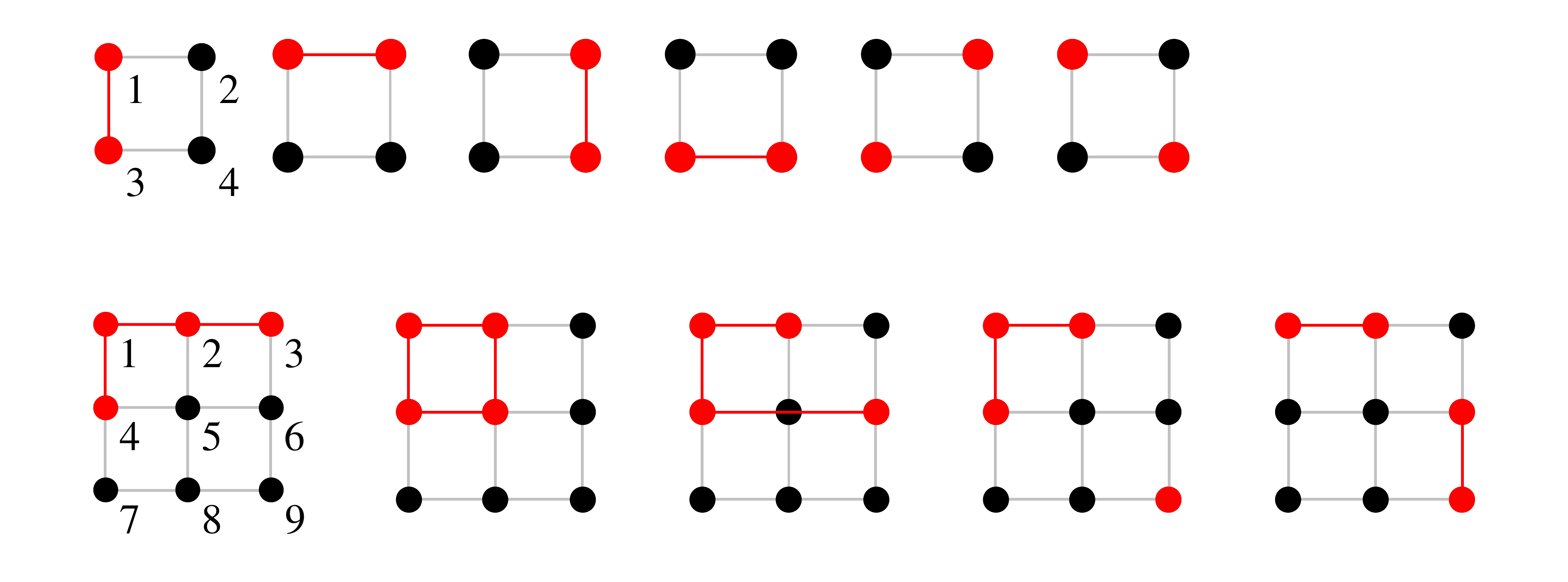} 
  	\caption{\emph{Top row:} All possible colorings of a plaquette with $N_{\rm exc}=2$ excitations, which can be devided in two equivalence classes $\theta_1$ (first four configurations) and $\theta_2$ (last two configurations). \emph{Bottom row:} Five equivalence classes $\theta_i$, $i=1..5$ for $N_{\rm exc}=4$ of $3 \times 3$ array. Only one representative of each class is shown. At the beginning of each row, we display the numbering of the array. Excitations are in red, ground state atoms in black.}
  	\label{fig:classes 3x3}
 	\end{figure}
\end{center}

Let us start with the following observation based on the exact diagonalization results of (\ref{eq:H spin int}) in the non-frustrated case $\lambda<0$ in the simplest non-trivial example, a plaquette (i.e. $2 \times 2$ array) with $N_{\rm exc}=2$ excitations. The vertices of the plaquette are labeled 1-4, see Fig. \ref{fig:classes 3x3}. The ground state can be written as
\beq
	\ket{\psi_{\rm GS}} = \frac{1}{\sqrt{2}} \left( \ket{\theta_1} + \ket{\theta_2} \right)
	\label{eq:GS plaquette}
\eeq
where
\beqa
	\ket{\theta_1} &=& \frac{1}{\sqrt{4}} \left ( \ket{1100}+\ket{1010}+\ket{0101}+\ket{0011} \right) \\
	\ket{\theta_2} &=& \frac{1}{\sqrt{2}} \left ( \ket{1001}+\ket{0110} \right),
	\label{eq:theta plaquette}
\eeqa
which we symbolically write as
\beq
	\ket{\theta_i} = \frac{1}{\sqrt{|\theta_i|}} \sum_{j \in \theta_i} \ket{s_j},
\eeq
where $\ket{s_j}$ is a specific spin configuration and $|\theta_i|$ the number of such configurations belonging to a given set $\theta_i$. This seemingly artificial decomposition of the ground state into $\ket{\theta_1}$ and $\ket{\theta_2}$ is in fact directly related to the coloring of a graph as we now discuss. 

Let us start by introducing the notions necessary for our considerations which we then demonstrate on specific examples of the ground state construction. To this end we follow closely the treatment presented in \cite{Balasubramanian_1985}.

~\\
\begin{itemize}
\item{Lets consider a set $S$ and a group $G$ acting on $S$ with ranks $|S|$ and $|G|$ respectively.}

\item{For $G$ a discrete group, each element $g \in G$ can be written as a product of $j-$cycles $x_j$, $g \rightarrow \left(x_1^{b_1} x_2^{b_2}...x_{|S|}^{b_{|S|}} \right)_g$, where $b_j$ counts how many $j-$cycles appear in the decomposition of $g$. The product $\left(x_1^{b_1} x_2^{b_2}...x_{|S|}^{b_{|S|}} \right)_g$ is thus a monomial representing the cycle structure of the element $g$}

\item{Lets consider $m$ colours $c_1,...,c_m$ such that a specific colour $c_j$ is assigned to each element of $S$}
\end{itemize}
~\\
\emph{Definition 1:} A \emph{colouring} $C$ is a specific configuration of colours on $S$.\\
~\\
For example, considering two colours (black and red), \CdisA is a possible colouring of a plaquette.\\
~\\
\emph{Definition 2:} An \emph{orbit} of a colouring $C$ is a set of all colourings produced by the action of the group $G$ on $C$, ${\rm orb}_G(C) = \{g(C), g \in G\}$\\
~\\
An orbit is thus an \emph{equivalence class} of all colourings belonging to the orbit.\\
~\\
\emph{Definition 3:} A \emph{stabilizer} of a colouring $C$ is a set of all group elements $g$ which leave $C$ invariant, ${\rm stab}_G(C) = \{g \in G, g(C)=C\}$\\
~\\
\emph{Definition 4:} A \emph{generating function} (or \emph{pattern inventory} or \emph{cycle index}) is a polynomial given by the sum of all monomials of elements of $G$ acting on $S$
\beq
	P_G(x_1,...x_{|S|}) = \frac{1}{|G|} \sum_{g \in G} \left(x_1^{b_1} x_2^{b_2}...x_{|S|}^{b_{|S|}} \right)_g
	\label{eq:generating function def}
\eeq
\\
~\\
With the definitions above we now introduce two theorems:\\
~\\
\emph{Theorem 1:} \emph{P\'{o}lya's enumeration theorem} \cite{Keller_2014}.  Let $\mathcal{C} = \{ C \}$ be a set of all colourings of $S$ using colours $c_1,...,c_m$. Let $G$ induce an equivalence relation on $\mathcal{C}$. Then 
\beq
	P_G(\sum_{i=1}^m c_i, \sum_{i=1}^m c_i^2,...,\sum_{i=1}^m c_i^{|S|} )
\eeq
is the generating function for the number of non-equivalent colorings of $S$ in $\mathcal{C}$. \\
~\\
\emph{Theorem 2:} \emph{Orbit-stabilizer theorem} \cite[ch.7]{Cameron_2008,Gallian_2006}.
\beq
	\left| {\rm orb}_G(C)\right| = \frac{|G|}{\left| {\rm stab}_G(C) \right|}
	\label{eq:orbit-stabilizer}
\eeq
\\
~\\
Equipped with the necessary notions, we return back to the example of the ground state (\ref{eq:GS plaquette}) \footnote{\label{fn:footnote1}The example of colouring of a plaquette is carried out in great detail in \cite{Keller_2014} and we refer the reader to this reference for further information.}. In order to find the structure of the ground state corresponding to a given excitation number sector $N_{\rm exc}$, we need to enumerate the number of the sets $\theta$ and how many elements belong to each of the set. Here, we are concerned only with two colours, say black and red, which correspond to spins in ground and excited state respectively. In other words, $\theta$ \emph{is} precisely an orbit and $|\theta|$ is thus given by the orbit-stabilizer theorem. We now demonstrate the use of the above theorems on our example of (\ref{eq:GS plaquette}). The generating functional (\ref{eq:generating function def}) of the $G_{2 \times 2} = \{\mathfrak{R},\mathfrak{C}\} \cup D_4 = D_4$ group of the plaquette reads \cite{endnote69}
%(see footnote \ref{fn:footnote1})
\beqa
	P_{G_{2 \times 2}} &=& \frac{1}{8}\left( x_1^4 + 2 x_1^2 x_2 + 3 x_2^2 +  2 x_4 \right) \nonumber \\
	 &=& b^4 + b^3 r + 2 b^2 r^2 + b r^3 + r^4.
	 \label{eq:generating function plaquette}
\eeqa
In the second line, we have used the P\'{o}lya's theorem, where we have substituted the black ($b$) and red ($r$) colours, $x_j = b^j + r^j$ for $j=1,2,4$. In our example of two excitations, i.e. the term with $r^2$ in (\ref{eq:generating function plaquette}), the numerical prefactor 2 means there are two equivalence classes $\theta_1, \theta_2$ of the colourings. These can be found explicitly and read
\beq
\begin{aligned}
	\theta_1 \equiv {\rm orb}_{G_{2 \times 2}} \left( \CparA \right) &= \left\{ \CparA, \CparB, \CparC, \CparD \right\} \\
	{\rm stab}_{G_{2 \times 2}} \left( \CparA \right) &= \left\{ e, h \right\} \\
	\theta_2 \equiv {\rm orb}_{G_{2 \times 2}} \left( \CdisA \right) &= \left\{ \CdisA, \CdisB \right\} \\
	{\rm stab}_{G_{2 \times 2}} \left( \CdisA \right) &= \left\{ e, r_{\pi}, p, n \right\},
\end{aligned}
\eeq
where $e$ stands for the identity element of the group $G$. Finally, one can verify that the above relations obey the orbit-stabilizer theorem (\ref{eq:orbit-stabilizer}) so that $|\theta_1| = 4$ and $|\theta_2|=2$ with the states written explicitly in (\ref{eq:theta plaquette}).\\
~\\
The above results can be generalized straightforwardly to larger arrays. In that case the ground state can be written as 
\beq
	\ket{\psi_{\rm GS}} = \sum_i \psi_i \ket{\theta_i},
	\label{eq:GS array}
\eeq
where the orbit states $\ket{\theta_i}$ are orthonormal, $\braket{\theta_i | \theta_j} = \delta_{ij}$. To give an explicit example going beyond the plaquette, we consider a $3 \times 3$ array and we choose $N_{\rm exc}=4$ sector. We find that the total of $\binom{9}{4}=126$ spin basis states form five equivalence classes depicted in Fig. \ref{fig:classes 3x3}

The Hamiltonian in the orbit states basis $\{\ket{\theta_1},\ket{\theta_2}\,\ket{\theta_3},\ket{\theta_4},\ket{\theta_5}\}$ reads
\beq
H_{\rm spin, int} = 
\left(
\begin{array}{ccccc}
 3 & 0 & 2 & 2 & 2 \\
 0 & 0 & 4 & 0 & 0 \\
 2 & 4 & 2 & 4 & 0 \\
 2 & 0 & 4 & 4 & 4 \\
 2 & 0 & 0 & 4 & 0
\end{array}
\right).
\label{eq:H 3x3}
\eeq
We note that $\bra{\theta_2} H_{\rm spin, int} \ket{\theta_2} = \bra{\theta_5} H_{\rm spin, int} \ket{\theta_5} = 0$. This can be easily understood when inspecting the structure of $\theta_2, \theta_5$ and realizing that the action of the Hamiltonian (\ref{eq:H spin int}) is to anihilate an excitation at a given site and create an excitation at another site. It can be seen from Fig. \ref{fig:classes 3x3} that such operations necessarily take a state belonging to $\theta_2$ or $\theta_5$ out of its equivalence class. For instance for the example of $\theta_2$ in Fig. \ref{fig:classes 3x3}, anihilating the excitation at position 5 and creating an excitation at position 6 results in the state $\theta_3$ shown and the corresponding non-zero matrix element $\bra{\theta_2} H_{\rm spin, int} \ket{\theta_3}$ in (\ref{eq:H 3x3}) (in fact an operation displacing \emph{two} excitations at once would be required for the state to remain in $\theta_2$).

While the above considerations offer a useful insight into the structure of the ground state, so far they do not constitute a clear computational advantage as we did not provide a prescription for obtaining the Hamiltonian (\ref{eq:H spin int}) in the $\{ \ket{\theta_i} \}$ basis. Such prescription is likely to be equivalent to finding the irreps of $G$ as discussed at the beginning of this section. We leave this investigation for future work and restrict ourselves only to exact diagonalization in the comparative QMC study of the correlations presented in the following section.

\subsection{Correlations}

\begin{center} 
\begin{figure}[h!]
\vspace{0cm}
\includegraphics[width=0.75\columnwidth]{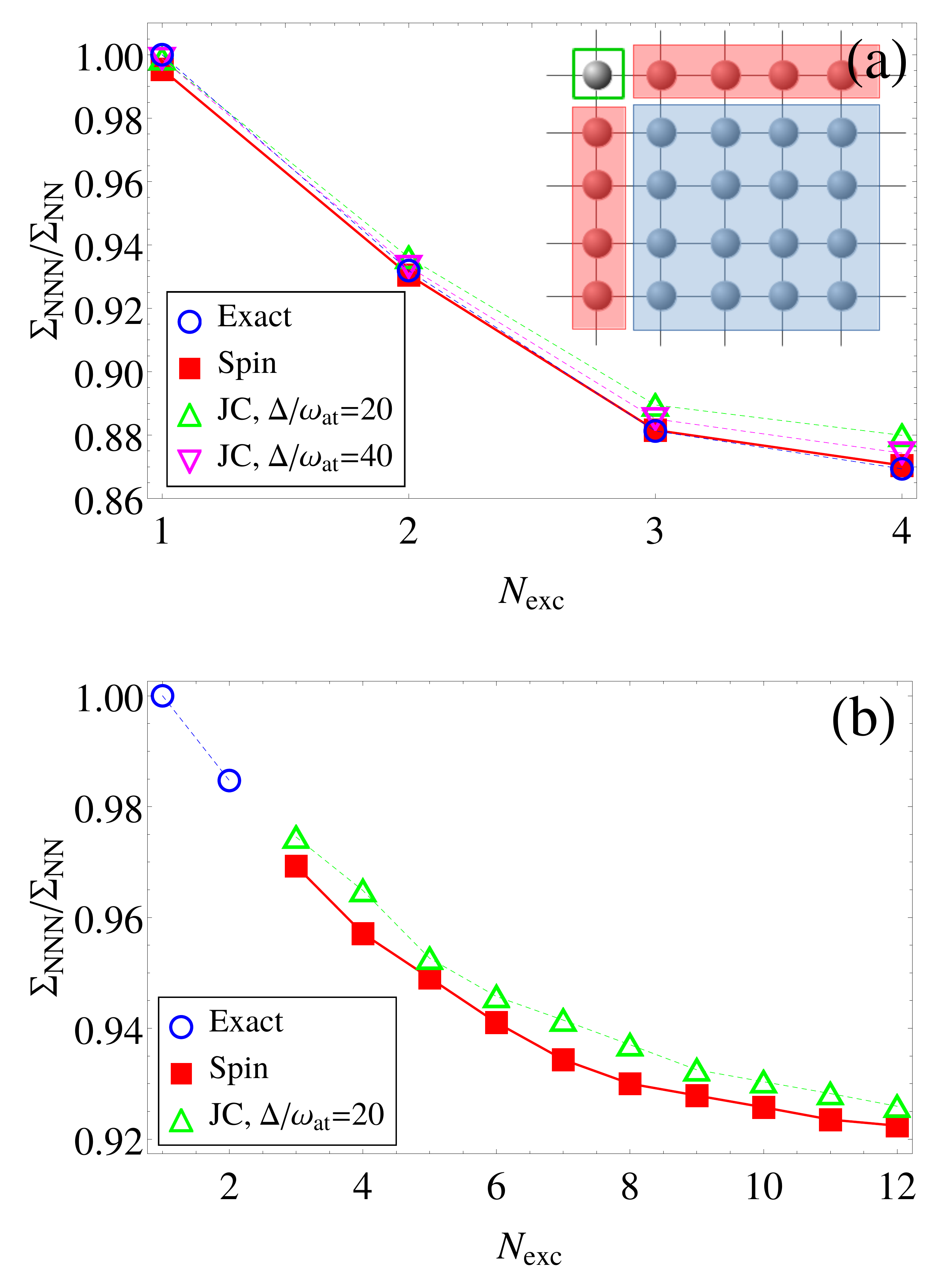}
%\vspace{-1.cm}
\caption{Correlations of the ground state of the spin model (\ref{eq:H spin main}) in (a) a $N=3 \times 3$ and (b) $N=5 \times 5$ array. The nearest-neighbour to next-to-nearest-neighbour ratio $\Sigma_{\rm NNN}/\Sigma_{\rm NN}$ of the connected correlation functions is plotted as a function of the number of spin excitations $N_{\rm exc}$. In (a) we find good agreement between the exact diagonalization results of the spin model (blue circles), the QMC simulation of the spin model (red squares) and the QMC simulation of the JC model (green and magenta triangles for $\Delta/\omega_{\rm at}=20$ and $40$ respectively). (b) Similar agreement is obtained for the $5 \times 5$ array. The inset in (a) is a schematic representation of the connectivity in the cavity array: NN is represented by the red-shaded (NNN by the blue-shaded) regions, see text for details.
} 
%\vspace{-2.cm}
\label{fig:correlations}
\end{figure}
\end{center}

We are now in position to study the correlation functions of the spin model. To this end we consider (connected) two-point correlations of the type $\braket{\sigma^+_{i\nu} \sigma^-_{j\mu}}$. Due to long (infinite) range connectivity along the rows and the columns, the system features only two length scales, the nearest-neighbours (NN, spins belonging to the same cavity mode) and next-to-nearest-neighbors (NNN) as there are at most two different cavity modes connecting any two spins of the array. We thus define two types of correlation functions, $\Sigma_{\rm NN} \equiv \{ \braket{ \sigma^+_{i\nu} \sigma^-_{i\mu}},\nu \neq \mu \} \cup \{ \braket{\sigma^+_{i\nu} \sigma^-_{j\nu}},i \neq j \}$ and $\Sigma_{\rm NNN} \equiv \{ \braket{ \sigma^+_{i\nu} \sigma^-_{j\mu}}, i \neq j,\nu \neq \mu \}$, where we have excluded self-correlations of the type $\braket{\sigma^+ \sigma^-} = \braket{ \ket{1}\bra{1}}$. This situation is schematically depicted in the inset of Fig. \ref{fig:correlations}a. Here, $\Sigma_{\rm NN}$ corresponds to correlations between the spin in the green box and the spins belonging to the same row and column (red-shaded region). Similarly, $\Sigma_{\rm NNN}$ corresponds to correlations between the spin in the green box and the spins belonging to the blue-shaded region. In Fig. \ref{fig:correlations} we plot the ratio $\Sigma_{\rm NNN}/\Sigma_{\rm NN}$ as a function of the number of spin excitations $N_{\rm exc}$ in an $N=3 \times 3$ (Fig. \ref{fig:correlations}a) and $N=5 \times 5$ array (Fig. \ref{fig:correlations}b). For the $3 \times 3$ array we find perfect agreement between the exact results obtained by exact diagonalization of the spin Hamiltonian (\ref{eq:H spin int}) in each excitation sector and the QMC simulation of that Hamiltonian \footnote{The QMC simulation is by construction performed in grand-canonical ensemble. The values of $\Sigma_{\rm NNN}/\Sigma_{\rm NN}$ were obtained by post-selecting on the results with integer number of excitations, $\sum_{i,\nu} \braket{n_{i \nu}}=N_{\rm exc}$.}. Moreover we find a good agreement also with the QMC simulation of the JC model (\ref{eq:H JC}), which is improving with increasing value of $\Delta/\omega_{\rm at}$ as it should. Similar agreement between the QMC simulations of the spin and the JC model is observed for the $5 \times 5$ array 
%(here, the exact diagonalization results were computed only for $N_{\rm exc}=1,2$).

%****************************************************************************************************************************************************************************************************************************
%****************************************************************************************************************************************************************************************************************************
\section{Towards simulation of frustrated spin systems in cavity arrays}
\label{sec:Frustration}

In Sec. \ref{sec:1D} we have shown, that in one dimension it is not possible to obtain the effective spin Hamiltonian (\ref{eq:H spin main}) with frustrated interactions $\lambda > 0$. The aim of this section is to show that this limitation can be circumvented in two dimensions by exploiting the properties of the $\delta \omega$ term (\ref{eq:delta omega spin}). 

In order to see this, we first perform a back-of-the-envelope estimation. The reason of the breakdown of the effective model in the frustrated case in one dimension is that as $\lambda$ is decreased, the term $\lambda a^\dag a$ in (\ref{eq:H spin 1D}) is decreasing in a way that the $n=0 \leftrightarrow n=\infty$ transition occurs before any spin excitation can appear. 

In order to simplify the analytical treatment, from now on we assume all the couplings along rows to be the same, $\lambda_a \equiv \lambda_i, \; \forall i$ and similarly for the columns, $\lambda_b \equiv \lambda_\nu, \; \forall \nu$. Assuming further that the row (column) cavity photon occupation numbers are $n_a$ ($n_b$), the expectation value of the $\delta \omega$ term (\ref{eq:delta omega spin}) becomes
\beq
	\braket{\delta \omega} = 2(\lambda_a n_a + \lambda_b n_b) + \sqrt{n_a n_b} (\lambda_a + \lambda_b) + (\lambda_a + \lambda_b)
	\label{eq:delta omega exp}
\eeq
It is apparent from the above expression that the magnitude of $\braket{\delta \omega}$ can be made small when the couplings along rows and columns have opposite signs, $\lambda_a = -\lambda_b$, so that the amplitude of each bracket in (\ref{eq:delta omega exp}) becomes significantly smaller than if we take both $\lambda$ with the same sign. We will thus consider a scenario with frustrated interactions along one direction (we choose the $a$ cavity modes) and non-frustrated one along the other ($b$ modes) and we paramterize the couplings as
\beq
  \lambda_a > 0, \;\;\; \lambda_b < 0, \;\;\; \lambda_b  = \eta \lambda_a.
  \label{eq:cond}
\eeq

In order to show that one can get a non-trivial frustrated--non-frustrated situation without breaking the effective spin description, we will use a self-consistency argument as follows. First we restore the free cavity fields terms we omitted in (\ref{eq:H spin main}) so that the effective spin Hamiltonian can be written as (see Appendix \ref{app:Adiabatic Elimination})
\beq
	H = H_{\rm ph} + H_{\rm spin,int}
\eeq
where 
\beq
	H_{\rm ph} = \Delta_a \sum_j a_j^\dag a_j + \Delta_b \sum_\mu b^\dag_\mu b + \sum_{j=1}^{L_x} \sum_{\mu=1}^{L_y} \delta \omega_{{\rm at},j\mu} \sigma^z_{j\mu}
	\label{eq:H phot}
\eeq
is now the photonic part (note that we have absorbed the $\delta H$ term (\ref{eq:delta H})) in $H_{\rm ph}$ and $H_{\rm spin,int}$ is given by (\ref{eq:H spin int}). Exploiting the fact that the $\sigma^z$ term is diagonal in eigenstate basis in all excitation sectors $N_{\rm exc}$ of the spin Hamiltonian (\ref{eq:H spin main}), we substitute the spin expectation values $\braket{\sigma^z}$ in (\ref{eq:H phot}) and subsequently diagonalize the photonic Hamiltonian, which is a straightforward exercise as it is quadratic in the photonic degrees of freedom. We then compare the values of the critical couplings at which a transition $N_{\rm exc} \rightarrow N_{\rm exc}+1$ occurs with that of $0 \rightarrow \infty$ photon number in analogy to the analysis in Sec. \ref{sec:1D}. We anticipate that a non-trivial frustrated regime can be always obtained by appropriate tuning of the system parameters and in particular its geometry. This is also the regime which fulfills the self-consistency criterion, namely taking $\braket{\delta \omega}=0$ in the spin model, using the corresponding solutions in the photonic Hamiltonian (\ref{eq:H phot}) and finding that its solutions again yield $\braket{\delta \omega}=0$.

\subsection{Diagonalization of the spin Hamiltonian}

In analogy to Sec. \ref{sec:Simulations} we seek to diagonalize the interaction part of the spin Hamiltonian (\ref{eq:H spin int})
\begin{equation}
	H=\lambda_a \sum_{j=1}^{L_x} \sum_{\mu \neq \nu=1}^{L_y} \sigma^+_{j \mu} \sigma^-_{j \nu} + \lambda_b \sum_{\alpha=1}^{L_y} \sum_{k \neq l=1}^{L_x} \sigma^+_{k \alpha} \sigma^-_{l \alpha} + {\rm h.c.}, \nonumber
	\label{eq:H spin int 2}
\end{equation}
where we have now explicitly written the summation limits. 
%In what follows, when we write explicitly the eigenstate of the spin Hamiltonian we use a simple letter to denote the combined position index, $p \equiv l\nu$ (i.e. a spin in $l$-th row and $\nu$-th column).

~\\
\noindent{\emph{0-excitation sector}} \\
Here the situation is trivial, the unique ground state being simply $\ket{\psi^0} = \ket{00..0}$, i.e. all spins down and correspondingly  $s^z_{l\mu} \equiv \bra{\psi^0} \sigma^z_{l\mu} \ket{\psi^0} = -1,\;\forall l,\mu$\\
~\\
\noindent{\emph{1-excitation sector}} \\
In the basis $\{ \ket{100..0}, \ket{010..0},...,\ket{000..1}\}$ of single particle states, the interaction Hamiltonian (\ref{eq:H spin int}) takes a simple form
\beq
	H_{1\rm exc} = 2 \lambda_a M_a \otimes \mathds{1}_b + 2 \lambda_b \mathds{1}_a \otimes M_b,
	\label{eq:H1p}
\eeq
where the $a$ ($b$) matrices have dimensions $L_x \times L_x$ ($L_y \times L_y$) respectively and the prefactor 2 comes from accounting twice for each spin configuration in (\ref{eq:H spin int}). $M$ are matrices with 1 everywhere except the diagonal, where it is 0, $M_{ij} = 1-\delta_{ij}$. The corresponding eigenvalues and multiplicities are
\begin{center}
\begin{tabular}{ |c| c| c| }
\hline
  matrix & eigenvalue & multiplicity \\ \hline
    \multirow{2}{*}{$M_a$} & -1 & $L_x - 1$ \\ \cline{2-3}
   & $L_x-1$ & 1 \\ \hline 
   \multirow{2}{*}{$M_b$} & -1 & $L_y-1$ \\ \cline{2-3}
    & $L_y-1$ & 1 \\
    \hline
\end{tabular}
\end{center}
\vspace*{-0.5cm}
\begin{equation}
\phantom{abc}
\label{eq:Eigs M}
\end{equation}
Since $\lambda_a>0$ and $\lambda_b <0$, the minimum energy is
\beq
	E^{1\rm exc}_{\rm min} = -2 \lambda_a + (L_y-1) 2 \lambda_b
\eeq
with multiplicity $L_x - 1$. The corresponding eigenvectors are
\beqa
	E_a = -1 &\rightarrow& \ket{v^j_a} = \frac{1}{\sqrt{2}} (\ket{1_1} - \ket{1_j}), \;\;\; j=2..L_x \\
	E_b = L_y-1 &\rightarrow& \ket{v_b} = \frac{1}{\sqrt{L_y}} \sum_{j=1}^{L_y} \ket{1_j},
\eeqa
where $\ket{1_j} \equiv \ket{0..01_j0..0}$. The ground state eigenvectors of $H_{1\rm exc}$ are then $\ket{\psi^0_j} = \ket{v_a^j} \otimes \ket{v_b}$ and the spin expectation values become
\beq
	\bra{\psi^0_j} \sigma^z_{l\mu} \ket{\psi^0_j} = \frac{1}{2L_y} - (1-\frac{1}{2 L_y}) = -1 + \frac{1}{L_y}
\eeq
if $\ket{\psi^0_j}$ contains the excitation at site $l\mu$ or -1 otherwise. We note that $s^z_{l\mu} \rightarrow -1$ in the thermodynamic limit as one would expect.

In summary, we have for the ground state energies
\begin{subequations}
\begin{align}
	E^{0\rm exc}_{\rm GS} &= \frac{\omega'_{\rm at}}{2} (-N) \\
	E^{1\rm exc}_{\rm GS} &= \frac{\omega'_{\rm at}}{2} (-N+2) -2 \lambda_a + (L_y-1) 2 \lambda_b,	
\end{align}
  \label{eq:EGS spin}
\end{subequations}
where
\beq
	\omega_{\rm at}' = \omega_{\rm at} + 2(\lambda_a + \lambda_b).
	\label{eq:omega prime}
\eeq

\subsection{Diagonalization of the photonic Hamiltonian}
We are now in position to diagonalize the photonic quadratic form (\ref{eq:H phot}). In analogy to the previous paragraph, we start our examination in the 0-excitation sector. Here the expectation values of the spin operator is simply $s^z \equiv s^z_{l\mu} = -1,\;\forall l,\mu$, so that the photonic Hamiltonian becomes
\beq
	\braket{H_{\rm ph}} = p^\dag M_{\rm ph} p,
\eeq
where we have introduced $p=(a_1,..,a_{L_y},b_1,..,b_{L_x})^T$. The matrix $M_{\rm ph}$ can be written as
\beq
	M_{\rm ph} = \begin{pmatrix}
		W_a & G \\
		G^T & W_b
	\end{pmatrix}
\eeq
with
\beqa
	(W_a)_{ij} &=& \delta_{ij} \left[ \omega_a + 2 \lambda_a \sum_{\nu=1}^{L_x} s^z_{i\nu} \right], \;\;\; i,j=1..L_y \\
	(W_b)_{ij} &=& \delta_{ij} \left[ \omega_b + 2 \lambda_b \sum_{l=1}^{L_y} s^z_{l j} \right], \;\;\; i,j=1..L_x \\
	G_{ij} &=& (\lambda_a + \lambda_b) s^z_{ij},\;\;\; i=1..L_y,j=1..L_x.
\eeqa
$M_{\rm ph}$ has the following eigenvalues and multiplicities

\begin{center}
\begin{tabular}{ |c l| c| }
\hline
 & eigenvalue & multiplicity \\ \hline
 $E_a^{\rm ph} \equiv$ & $\Delta_a + 2 \lambda_a L_x s^z$ & $L_y-1$ \\
 $E_b^{\rm ph} \equiv$ &  $\Delta_b + 2 \lambda_b L_y s^z$ & $L_x-1$ \\
 $E_+^{\rm ph} \equiv$ & $\frac{1}{2}(\epsilon + \xi)$ & 1 \\   
 $E_-^{\rm ph} \equiv$ & $\frac{1}{2}(\epsilon - \xi)$ & 1 \\
    \hline
\end{tabular}
\label{tab:Eigs ph}
\end{center}
\vspace*{-0.5cm}
\begin{equation}
\phantom{abc}
\label{eq:Eigs ph}
\end{equation}
where
\begin{subequations}
\begin{align}
	\epsilon &= \Delta_a + \Delta_b + 2s^z(\lambda_a L_x + \lambda_b L_y) \\
	\xi &= \sqrt{(\Delta_b-\Delta_a + 2s^z(\lambda_a L_x - \lambda_b L_y))^2 + 4 L_x L_y (s^z)^2(\lambda_a+\lambda_b)^2}.
\end{align}
\end{subequations}

\subsection{Validity and breakdown of the effective spin model}
First we note, that due to (\ref{eq:cond}), $\Delta_b$ is not independent and can be expressed as
\beq
	\Delta_b = \frac{\Delta_a - \omega_{\rm at}}{\eta} + \omega_{\rm at}
\eeq
(here $\omega_{\rm at}$ is the bare atomic frequency, not $\omega_{\rm at}'$). Motivated by the condition (\ref{eq:cond elimination}) needed for the spin model (\ref{eq:H spin main}) to be valid, we define what we call the \emph{quality factor} of the approximation as
\beq
	Q = {\rm min} \left( \frac{|\Delta_a - \omega_{\rm at}|}{g_c^{\rm spin}}, \frac{|\Delta_b - \omega_{\rm at}|}{g_c^{\rm spin}} \right),
\eeq
where $g_c^{\rm spin}$ is given by the critical value of $\lambda_{a}$, $g_c^{\rm spin} = \sqrt{-2 \lambda_{a,{\rm crit}} (\Delta_a - \omega_{\rm at})}$, cf. below.

We start by determining the critical value of $\lambda_a$ at which a transition from 0- to 1-excitation sector occurs in the spin model. This can be simply obtained from the condition $E^{0\rm exc}_{\rm GS}=E^{1\rm exc}_{\rm GS}$ and with the help of (\ref{eq:EGS spin}) we get
\beq
	\lambda_{a,c}^{\rm spin} = -\frac{\omega_{\rm at}}{2\eta L_y}.
\eeq

Next, the breakdown of the effective model is indicated when \emph{any} of the photonic eigenvalues (\ref{eq:Eigs ph}) becomes negative, corresponding to infinitely many photons in the ground state. Substituting $s^z=-1$ in (\ref{eq:Eigs ph}), the only two candidates for the minimum eigenvalue are $E_a^{\rm ph}$ and $E_-^{\rm ph}$ (we recall that $\lambda_a>0,\lambda_b<0$). The value of $\lambda_a$ where $E^{\rm ph}$ becomes negative is determined as
\beq
	\lambda_{a,c}^{\rm ph} \equiv {\rm min} \left( \lambda_a\,:\, E_a^{\rm ph} (\lambda_a) = 0, E_-^{\rm ph} (\lambda_a) = 0  \right),
\eeq
where only the positive branch of $\lambda_a$ in the solutions of $E_-^{\rm ph} (\lambda_a) = 0$ is considered. 

The criterion of having a valid and non-trivial regime in the effective spin Hamiltonian (i.e. finite number of photons and non-zero spin excitations) thus translates into 
\beq
	R \equiv \frac{\lambda_{a,c}^{\rm ph}}{\lambda_{a,c}^{\rm spin}} > 1 \;\; \wedge \;\; Q \gg 1
	\label{eq:cond breakdown}
\eeq

\begin{center}
	\begin{figure}[t!p]
  	\includegraphics[width=13cm]{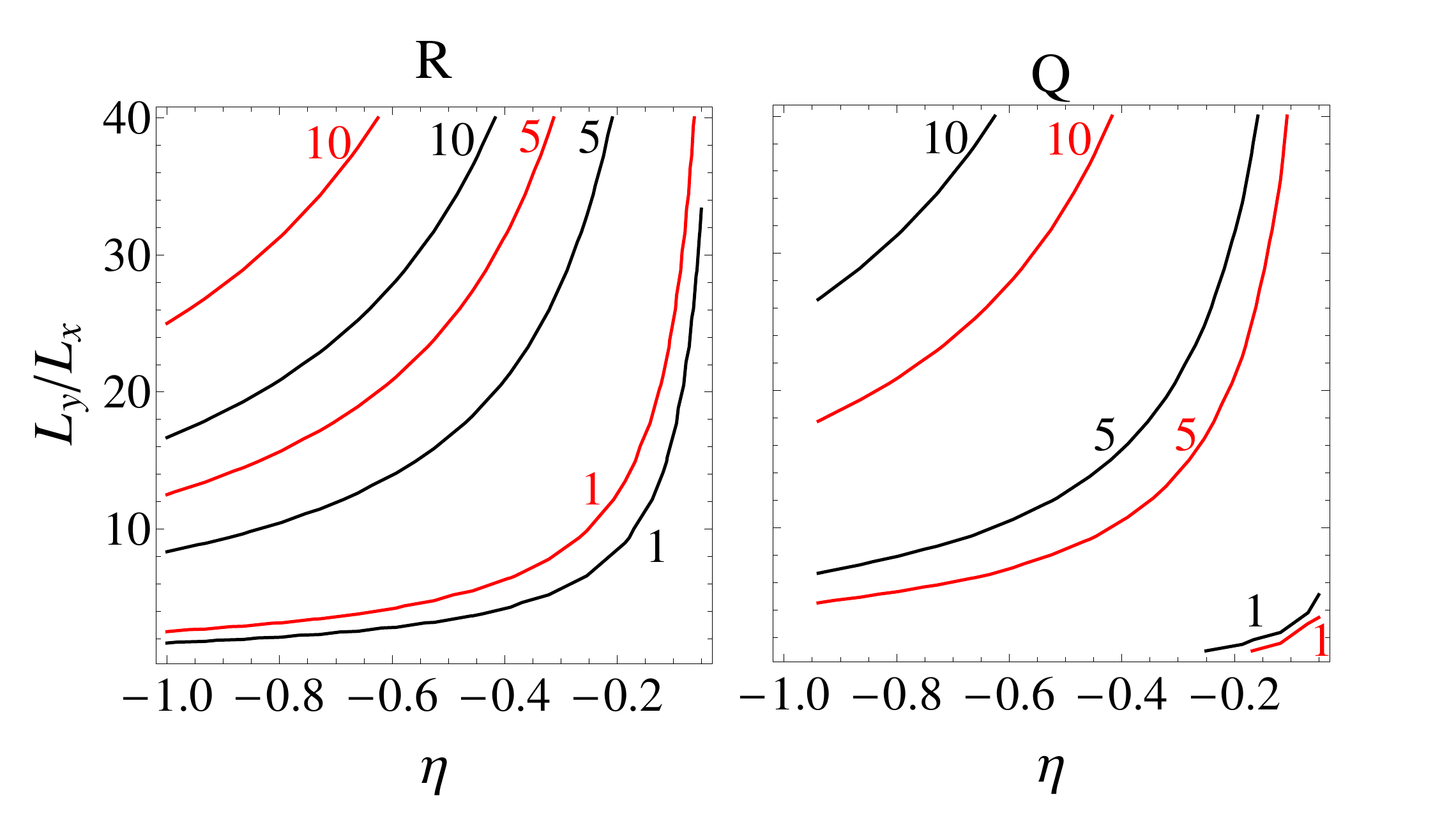} 
  	\caption{Region of validity of the effective spin Hamiltonian in the single excitation sector. Left (right) pane shows contour plot of the parameter $R$ ($Q$) in the $L_y/L_x - \eta$ plane. The effective spin model is valid provided $R>1$ and $Q \gg 1$, see condition (\ref{eq:cond breakdown}). The red (black) contour lines correspond to $\Delta_a/\omega_{\rm at} = 0.4$ (0.6) respectively. Parameters used: $L_x=10$.}
  	\label{fig:breakdown}
 	\end{figure}
\end{center}

In Fig. \ref{fig:breakdown} we plot the contours of constant $R$ (left pane) and $Q$ (right pane) respectively in the $\eta$--$L_x/L_y$ plane. It is apparent from the figure that increasing both $L_y/L_x$ and $|\eta|$ leads to a larger critical ratio $R$, i.e. we can ensure the presence of the non-trivial region by tuning these parameters. Additionally, one can show analytically that
\beqa
	\lim_{L_y \rightarrow \infty} R &=& \infty \\
    \lim_{\eta \rightarrow \infty}R &=& R_{\rm asymptotic}
\eeqa
when keeping all the other parameters fixed, as expected from the contours in Fig. \ref{fig:breakdown} (here $R_{\rm asymptotic}$ is some finite asymptotic value).

So far we have concetrated only on the simple 0 and 1-excitation sectors of the spin Hamiltonian. Clearly, for the interactions to become relevant one is interested in sectors with larger number of excitations. To this end one could extend the above analysis to the $N_{\rm exc} \rightarrow N_{\rm exc}+1$ particle transitions for $N_{\rm exc} \geq 1$ and for each of the transitions evaluate the critical ratio $R$. The fully analytical approach is unfortunatelly obscured by the fact that for  particle sectors $N_{\rm exc} >1$, the spin interaction matrix (\ref{eq:H spin int}) does not take the simple structure of (\ref{eq:H1p}) and the evaluation of eigenvalues thus amounts to solving higher order polynomials which in general requires a numerical approach. However, the arbitrarily large values of $R$ for the $0 \rightarrow 1$ spin excitation transition are suggestive and it is likely that also higher excitation number sectors can be reached without breaking the validity of the spin Hamiltonian.

To recap, we have shown that the spin model (\ref{eq:H spin main}) with both frustrated and non-frustrated interactions can emerge as an effective description of the parent JC Hamiltonian (\ref{eq:H JC}). This can be achieved when considering an elongated geometry of the square array, $L_y \gg L_x$. On one hand this circumvents the limitations related to realizing effective spin Hamiltonians with frustrated interactions using optical setups governed by JC Hamiltonians \cite{Rotondo_2015}. Finally, we note that one can simulate the parent JC model in the regime where it yields the effective spin model using QMC avoiding thus a sign problem of the spin model which opens up an interesting perspective on the QMC simulations of Hamiltonians with sign problem.

%****************************************************************************************************************************************************************************************************************************
%****************************************************************************************************************************************************************************************************************************
\section{Conclusions and Outlook}
\label{sec:Conclusions}

In this work we have analysed the ground states of a cavity array where each intersection of cavity modes is occupied by a single atom. We have shown that the system's description in terms of the JC model leads to an effective description - in a suitable parameter regime, where the cavity modes can be adiabatically eliminated - in terms of a two-component spin model. In one dimension, we have provided exact solution of the spin model demonstrating explicitly the need of higher dimensions in order to obtain frustrated spin-spin interactions. Using large-scale QMC simulation of the JC model we have performed a quantitative comparison between the parent JC and the emerging spin model. Specifically, in two dimensions, we have studied the superradiant phase transition and the properties of two-point correlation functions in the cavity array and we have described the graph-theoretical structure of the ground states of the spin model. In all cases we found a firm agreement between the two models in the regime of validity of the approximations used. Finally, we have outlined the possibility, by exploiting the non-linearities of the effective spin model, of studying frustrated spin Hamiltonians using two-dimensional cavity arrays.

In conclusion, the theoretical framework and numerical tools established in this work open ways to address the cavity QED physics in a quantitative way beyond the traditional mean-field or perturbative treatments. The present developments can be exploited in various scenarios, such as the study of the ground state properties in different lattice geometries and dimensions.

\section{Acknowledgments}

J. M. would like to thank Matteo Marcuzzi for useful discussions. The research leading to these results has received funding from the European Research Council under the European Union's Seventh Framework Programme (FP/2007-2013) / ERC Grant Agreement No. 335266 (ESCQUMA) and the EU-FET grants HAIRS 612862 and QuILMI 295293.

%%%%%%%%%%%%%%%%%%%%%%%%%%%%%%%%%%%%%%%%%%%%%%%%%%%%%%%%%%%%%%%
%\bibliographystyle{apsrev4-1}
%\bibliographystyle{prsty}
%\bibliography{"G:/Dokumenty/PHYS/Work/Many Body/Tex/cQED for MC/cavity_new/cavity_paper/cqed_bib"}
%\bibliography{"G:/Dokumenty/PHYS/Work/MANYBO~1/Tex/CQEDFO~1/cavity_new/cavity_paper/cqed_bib"}

%\bibliography{../cqed_bib}
%\bibliographystyle{apsrev4-1}
%\bibliographystyle{../prsty}
%\bibliographystyle{/media/ppzjm2/WD_G/Dokumenty/PHYS/Work/prsty}
%\bibliographystyle{G:/Dokumenty/PHYS/Work/prsty}

%\begin{widetext}

%****************************************************************************************************************************************************************************************************************************
%****************************************************************************************************************************************************************************************************************************
\appendix

\section{Adiabatic elimination of the cavity fields}
\label{app:Adiabatic Elimination}

We provide the details of the adiabatic elimination of the excited state $\ket{e}$ in our previous publication \cite{Minar_2016_pub} (note that here we use the basis $\{ \ket{e}, \ket{1}, \ket{0} \}$ instead of $\{ \ket{e}, \ket{s}, \ket{g} \}$ in \cite{Minar_2016_pub}). The resulting Hamiltonian reads
\beq
  H = \sum_i \Delta_i a^\dag_i a_i + \sum_\nu \Delta_\nu b^\dag_\nu b_\nu + \sum_{i,\nu} \left( \frac{\omega_{{\rm at},i\nu}}{2} + \delta \omega_{{\rm at},i\nu}  \right) \sigma^z_{i\nu} + g\left( \sigma^+_{i\nu} (a_i + b_\nu) + {\rm h.c.} \right)+F_{i \nu},
  \label{eq:H el}
\eeq
where
\beqa
	\omega_{a,i\nu} &=& - \frac{\Omega^2}{\Delta_e} \nonumber \\
	\delta \omega_{a,i\nu} &=& - \frac{g_0^2 \left( a_{i}^\dag + b_{\nu}^\dag  \right) \left( a_{i} + b_{\nu}  \right)}{2 \Delta_e} \nonumber \\
	g &=& -\frac{g_0 \Omega}{\Delta_e} \nonumber \\
	F_{i \nu} &=& -\frac{1}{2 \Delta_e} \left( \Omega^2 + g_0^2 \left( a_{i}^\dag + b_{\nu}^\dag  \right) \left( a_{i} + b_{\nu}  \right) \right)
	\label{eq:pars JC}
\eeqa
with $\Delta_x = \omega_x - (\omega_1 + \omega_T)$, $x=i,\nu,e$ and $g_0$ the coupling strength of the $\ket{1}-\ket{e}$ transition. We take $\Omega$ and $g_0$ to be real and positive throughout the article. Here we have set $\omega_{\rm aux} = \omega_1 + \omega_T$ in \cite{Minar_2016_pub} and carried out the adiabatic elimination under the usual condition $|\Delta_e| \gg |\Omega|,|g_0|$. 

In this article we focus on the regime, where $|\omega_{\rm at}| \gg |\braket{\delta \omega}|$. This can be in principle always achieved in the limit 
\beq
	\Omega \gg g_0.
	\label{eq:cond 1}
\eeq
Neglecting the $\delta \omega$ term (and consequently the $F$ term), the Hamiltonian (\ref{eq:H el}) simplifies to (\ref{eq:H JC}),
\beq
H_{\rm JC} = \sum_i\Delta_ia^{\dag}_ia_i+\sum_{\nu}\Delta_{\nu}b^{\dag}_{\nu}b_{\nu}+\sum_{i,\nu}\frac{\omega_{{\rm at},i\nu}}{2}\sigma^z_{i\nu}+ g \sum_{i,\nu}\left( \sigma^+_{i\nu}(a_{i} + b_{\nu}) + (a^{\dag}_{i} + b^\dag_{\nu}) \sigma^-_{i\nu} \right),
\eeq
which is the usual Jaynes-Cummings Hamiltonian. 

Next we proceed with the derivation of the Hamiltonian Eq. (\ref{eq:H spin main}). From here on we take the bare atomic frequencies to be equal for all atoms, $\omega_{{\rm at},i\nu} = \omega_{\rm at}, \; \forall i,\nu$. We find the form of the effective spin Hamiltonian after further elimination of the cavity fields. Working in the perturbative regime 
\beq
	\epsilon_{i\nu} = |g /(\omega_{{\rm at}} - \Delta_{i(\nu)})| \ll 1,\; \forall i,\nu,
	\label{eq:cond 2}
\eeq
one can eliminate the cavity fields iteratively by means of unitary transformation to arbitrary order in $\epsilon_{i \nu}$ \cite{Tannoudji_1998}
\beq
  H_{\rm spin} = {\rm e}^S H_{\rm JC} {\rm e}^{-S} = H^{\rm JC}_0 + H^{\rm JC}_{\rm int} + \comm{S}{H^{\rm JC}_0} + \comm{S}{H^{\rm JC}_{\rm int}} + \frac{1}{2}\comm{S}{\comm{S}{H^{\rm JC}_0}} + ...,
\eeq
where $H^{\rm JC}_0$, $H^{\rm JC}_{\rm int}$ stand for the free and interaction part of the JC Hamiltonian respectively. To first order in $\epsilon_{i\nu}$, the antihermitian matrix $S$ reads 
\beq
  S = \frac{g}{\delta_i} \left( a^\dag_i \sigma^-_{i\nu} - \sigma^+_{i\nu} a_i  \right) + \frac{g}{\delta_{\nu}} \left( b^\dag_{\nu} \sigma^-_{i\nu} - \sigma^+_{i\nu} b_{\nu}  \right),
\eeq
where $\delta_{i(\nu)} = \Delta_{i(\nu)} - \omega_{\rm at}$. The effective spin Hamiltonian becomes
\beqa
	H_{\rm spin} &=& H^{\rm JC}_0 + \frac{1}{2} \comm{S}{H^{\rm JC}_{\rm int}} \nonumber \\
			&=& \Delta_i a^\dag_i a_i + \Delta_\nu b^\dag_\nu b_\nu + \left( \frac{\omega_{\rm at}}{2} + \delta \omega^{\rm spin}_{{\rm at},i\nu} \right) \sigma^z_{i\nu} + \lambda_i \left( \sigma^-_{i\nu} \sigma^+_{i \mu} + \sigma^+_{i\nu} \sigma^-_{i\mu} \right) + \lambda_\nu \left( \sigma^-_{i\nu} \sigma^+_{j \nu} + \sigma^+_{i\nu} \sigma^-_{j\nu} \right), \nonumber \\
	\label{eq:H spin tot}
\eeqa
where
\beqa
  \delta \omega^{\rm spin}_{{\rm at},i\nu} &=& \lambda_i \left( 2 a^\dag_i a_i + 1 + a^\dag_i b_\nu + b^\dag_\nu a_i \right) + \lambda_\nu \left( 2 b^\dag_\nu b_\nu + 1 + a^\dag_i b_\nu + b^\dag_\nu a_i \right) \nonumber \\
  \lambda_i &=& -\frac{g^2}{2 \delta_i} \nonumber \\
  \lambda_\nu &=& -\frac{g^2}{2 \delta_\nu}.
  \label{eq:pars spin}
\eeqa

%****************************************************************************************************************************************************************************************************************************
%****************************************************************************************************************************************************************************************************************************
\section{Mean-field solution of the JC model}
\label{app:MF}

The Hamiltonian of the system in the $2D$ case is the following:
\begin{equation}
H_{JC} = \Delta \left(\sum_i a^{\dagger}_i a_i +\sum_{\nu} b^{\dagger}_{\nu} b_{\nu} \right) + \frac{\omega_{\rm at}}{2}\sum_{i \nu} \sigma_{i\nu}^z + g \sum_{i\nu}\left(a^{\dagger}_i \sigma^-_{i\nu}+a_i \sigma^+_{i\nu}\right) + + g \sum_{i\nu}\left(b^{\dagger}_{\nu} \sigma^-_{i\nu}+b_{\nu} \sigma^+_{i\nu}\right)\,.
\end{equation}

We notice that the number of two-level atoms is $N = L_x \times L_y$ whereas the number of electromagnetic modes is $N_{\rm em}= L_x + L_y$. This means that for a large $2D$ array $N \gg N_{\rm em}$ and we can apply the standard mean field techniques originally introduced in refs. \cite{Wang_1973,Hepp_1973}. Since we are interested in the zero temperature limit, we restrict our analysis to this particular case, where the calculation amounts to average the full Hamiltonian over a set of photonic coherent states:
\begin{equation}
\ket{\alpha}_i = e^{-\frac{|\alpha_i|^2}{2}} e^{\alpha_i a_i^{\dagger}}\ket{0}\,, \qquad \ket{\beta}_{\nu} = e^{-\frac{|\beta_{\nu}|^2}{2}} e^{\beta_{\nu} b_{\nu}^{\dagger}}\ket{0}
\end{equation}
and to minimize the resulting non-interacting problem with respect to the set of variational complex variables $(\alpha_i,\beta_{\nu})$. By symmetry, the ground state solutions must be of the form $\alpha_i = \beta_{\nu} = \alpha$ and thus the partial integration over the photonic degrees of freedom gives the following function of the spins only:
\begin{equation}
H_{\rm JC}^{\rm MF} = \Delta (L_x + L_y) |\alpha|^2 + \frac{\omega_{\rm at}}{2} \sum_{i\nu}\sigma^z_{i\nu} + 2 g \left(\alpha \sum_{i\nu} \sigma^+_{i\nu} + \alpha^{\ast} \sum_{i\nu} \sigma^-_{i\nu} \right)\,,
\end{equation}
which can be easily diagonalized in every atomic subspace, giving as a result for the energy of the ground state:
\begin{equation}
E_{\rm GS} = \Delta \left( L_x + L_y \right) |\alpha|^2 - L_x L_y \sqrt{\left(\frac{\omega_{\rm at}}{2}\right)^2 + 4 g^2 |\alpha|^2}\,.
\end{equation}
The minimization of $E_{\rm GS}$ allows us to appreciate two different phases of the system: (i) a phase where the stable solution is $|\alpha|^2 = 0$, which physically corresponds to a zero macroscopic number of atomic excitations in the system (and also to a zero macroscopic number of photons in the cavities, since this number is proportional to $|\alpha|^2$). (ii) A superradiant phase where the stable solution is:
\begin{equation}
|\alpha_s|^2 = \left(\frac{L_x L_y g}{\Delta (L_x + L_y)}\right)^2 - \left(\frac{\omega_{\rm at}}{4g}\right)^2\,.
\end{equation} 
This solution is stable above a critical coupling which is given by:
\begin{equation}
g_c^{\rm MF} = \sqrt{\frac{(L_x+L_y)\omega_{\rm at} \Delta}{4 L_x L_y}}\,,
\end{equation} 
and physically represents the macroscopic number of photons in a given cavity of the array system. In the superradiant regime the atomic ground state of the system is:
\begin{equation}
\ket{\rm GS} = \bigotimes_{i,\nu=1}^{L_x,L_y} \ket{\rm GS}_{i\nu}\,, \qquad \ket{\rm GS}_{i\nu} = \frac{1}{\sqrt{\gamma^2 +1}} \left(\gamma \ket{+}_{i\nu} + \ket{-}_{i\nu}\right)\,,
\end{equation}
where $\ket{\pm}_{i\nu}$ are the eigenstates of $\sigma^z_{i\nu}$ with eigenvalues $\pm 1$ and 
\begin{equation}
\gamma = \frac{\frac{\omega_{\rm at}}{2}-\sqrt{\left(\frac{\omega_{\rm at}}{2}\right)^2 + 4g^2 |\alpha_s|^2}}{2g |\alpha_s|}\,,
\end{equation}
This expression can be used to evaluate analitycally the macroscopic number of atomic excitations:
\begin{equation}
\bra{\rm GS}\frac{1}{2}\left( L_x L_y + \sum_{i\nu} \sigma^z_{i\nu}\right) \ket{\rm GS} = \frac{1}{2}L_x L_y \left(1- \left(\frac{g_c}{g}\right)^2\right)
\end{equation}

%\begin{wrapfigure}{L}{0.4\textwidth}
%\vspace{-0.2cm}
% \centering
 %\includegraphics[angle=270, width=7cm]{conf_worm.pdf}
%\vspace{-0.3cm}
 %\caption{Example of a typical configuration. The red disconnected worldline is the \textbf{worm}.}
 %\end{wrapfigure}

%\textbf{Bibliography}
%\begin{enumerate}
%\item[1.] N. V. Prokof'ev, B. V. Svistunov, I. S. Tupitsyn, Phys. Lett. A 238, 253 (1998); Sov. Phys. JETP 87, 310 (1998).
%\item[2.] D. M. Ceperley and E. L. Pollock, Phys. Rev. B 36, 8343 (1987).
%\item[3.]  \c{S}. G. S\"{o}yler, B. Capogrosso-Sansone, N. V. Prokof'ev, B. V. Svistunov, New J. Phys. 11, 073036 (2009).
%\end{enumerate}

%\end{widetext}

\end{document}